\documentclass[showpacs,preprintnumbers,amsmath,amssymb]{revtex4}
\usepackage[latin1]{inputenc}
\usepackage[english]{babel}
\usepackage{graphicx}
\usepackage{amsfonts}
\usepackage{lscape}
\DeclareMathOperator{\arctanh}{arctanh}
\begin{document}
\title{Phase space of static wormholes sustained by an isotropic perfect fluid}
\author{Stéphane Fay\footnote{steph.fay@gmail.com}}
\affiliation{
Palais de la Découverte\\
Astronomy Department\\
Avenue Franklin Roosevelt\\
75008 Paris\\
France}
\begin{abstract}
A phase space is built that allows to study, classify and compare easily large classes of static spherically symmetric wormholes solutions, sustained by an isotropic perfect fluid in General Relativity. We determine the possible locations of equilibrium points, throats and curvature singularities in this phase space. Throats locations show that the spatial variation of the gravitational redshift at the throat of a static spherically symmetric wormhole sustained by an isotropic perfect fluid is always diverging, generalising the result that there is no such wormhole with zero-tidal force. Several specific static spherically symmetric wormholes models are studied. A vanishing density model leads to an exact solution of the field equation allowing to test our dynamical system formalism. It also shows how to extend it to the description of static black holes. Hence, the trajectory of the Schwarzschild black hole is determined. The static spherically symmetric wormhole solutions of several usual isotropic dark energy (generalised Chaplygin gas, constant, linear and Chevallier-Polarski-Linder equations of state) and dark matter (Navarro-Frenk-White profile) models are considered. They show various behaviours far from the throat: singularities, spatial flatness, cyclic behaviours, etc. None of them is asymptotically Minkowski flat. This discards the natural formation of static spherically symmetric and isotropic wormholes from these dark fluids. Last we consider a toy model of an asymptotically Minkowski flat wormhole that is a counterexample to a recent theorem claiming that a static wormhole sustained by an isotropic fluid cannot be asymptotically flat on both sides of its throat.
\end{abstract}
\pacs{95.30.Sf, 98.80.Jk}
\maketitle
%------------------------------------------------------------------------------%
\section{Introduction}\label{s0}
In this paper, we define a phase space to study, classify and compare large classes of static spherically symmetric wormholes solutions sustained by a perfect fluid in General Relativity and derive various physical results.\\
Such wormholes are defined by two free functions. We choose the first one by relating the radial pressure $p$ to the tangential pressure $\tau$ such as $p=-\tau$. The perfect fluid is then isotropic as a cosmological fluid (other choices of perfect fluid are possible with $p\not = -\tau$ as in \cite{Kuh17,Sus05,Lob06,Kuh16}). Such fluids are interesting for several reasons. In particular, they are generally (not always) physically motivated and constrained by observations such as the supernovae and the CMB. Another reason is the possibility that a ghost dark energy could both rules the Universe expansion and forms static spherically symmetric wormholes naturally, a possibility that we discard in this paper for some of the most common dark fluids used in the literature. The second free function can be the wormhole energy density, its pressure, its spatial shape, its gravitational redshift, etc. We let it free when studying general properties of the phase space and we specify it when examining some particular wormhole models.\\
Dynamical system analysis is often used in cosmology\cite{Wai05}. If it is not new for inhomogeneous spherical symmetric solutions, few of them have been studied that way. See for instance \cite{Mig89,Mig92,Pol95,Gan15,Cru17,Sus08,MonZan09}. Due to the complexity of the field equations, it is generally difficult to obtain exact wormhole solutions or study extensively some numerical ones. Some wormhole models are easier to analyse than others, as the ones with zero tidal-force contrary to those for which the form of the exotic fluid is specified (for instance by its equation of state). The dynamical system analysis is then particularly useful. It allows representing in a finite phase space, with an appropriate choice of normalised variables, all the solutions of a first order equations system describing a static spherically symmetric wormhole without having to solve them. Beyond the mathematical analysis, classification and comparison of wormhole models in a common framework, the dynamical system formalism also allows to get general physical results thanks to the determination of specific points in the phase space that can be common to any model. Among them, the equilibrium points representing transient or asymptotically behaviours like the Minkowski solutions, but also some non equilibrium points like those standing for the wormhole throats.\\
To find these points, we need to define a finite phase space by a set of normalised variables independent from the choice of the above mentioned free function and to rewrite the field equations. Then, in this general framework, we look for all the possible locations of throats, equilibrium points and curvature singularities in the phase space. To check the presence of singularities, we also calculate the geodesic equation for spacetimes associated to a phase space trajectory. We note that the points of the phase space corresponding to throats are not equilibrium points despite they are common to all phase space trajectories of wormhole models. After these mathematical general properties, we study several static spherically symmetric wormhole models in this unified framework and get some physical results.\\
A first model considers a vanishing density. Its exact solution is given. It allows checking the consistency of our dynamical system of equations with the original wormhole field equations system\cite{MorTho88}. From a physical viewpoint, it also shows how our static wormhole phase space can be used to represent static black holes as well although we did not pursue further in this paper dedicated to wormholes. We just determine the trajectory of the Schwarzschild black hole. Other models we consider are some usual cosmological dark fluids as some dark energy equations of state (generalised Chaplygin gas\cite{Ben02}, constant\cite{Rah07}, linear redshift parameterisation\cite{Hut01, Wel02}, Chevallier-Polarski-Linder parameterisation\cite{Che01,Lin03}) and a Navarro-Frenk-White\cite{Nav96} profile dark matter distribution. We classify their wormhole solutions and show that none of them is asymptotically Minkowski flat that disagrees with their natural formation from these dark fluids. Last, we consider a toy model of static isotropic wormhole that can be asymptotically Minkowski flat. We show that it is a counter-example to a recent theorem claiming that a static wormhole sustained by an isotropic fluid cannot be asymptotically flat on both sides of its throat.\\
Plan of the paper is as follows. In section \ref{s1}, we write the static spherically symmetric wormhole field equations as an autonomous system of normalised variables. We relate some of its phase space geometrical properties to some wormholes properties. In section \ref{s2}, we determine the possible locations of equilibrium points in the phase space. In section \ref{s3}, we do the same for the possible locations of curvature singularities. In section \ref{s4}, we study several static spherically symmetric wormhole models phase space and find the above mentioned physical results. We conclude in section \ref{s5}.
%------------------------------------------------------------------------------%
\section{Phase space}\label{s1}
In the first subsection, we cast the wormhole field equations \cite{MorTho88} into an autonomous dynamical system. We show how some large classes of wormholes can be studied in this unified framework. In the second subsection, we relate some phase space geometrical properties to some physical wormholes properties.
%------------------------------------------------------------------------------%
\subsection{Dynamical system}\label{s11}
The metric for a static spherically symmetric wormhole writes
\begin{equation}\label{met}
ds^2=-e^{2\Phi}dt^2+\frac{dr^2}{1-b/r}+r^2(d\theta^2+\sin^2\theta d\phi^2)
\end{equation}
$b(r)$ determines its spatial shape and $\Phi(r)$, its gravitational redshift. Considering General Relativity and a stress energy tensor for a perfect fluid defined by a density $\rho$, a pressure $p$ and a tension per unit area $\tau$, the field equations are\cite{MorTho88}
\begin{equation}\label{eq1A}
\dot b=8\pi G c^{-2} r^2 \rho
\end{equation}
\begin{equation}\label{eq2A}
\dot\Phi=(-8\pi G c^{-4}\tau r^3+b)/\left[2r(r-b)\right]
\end{equation}
\begin{equation}\label{eq3A}
\dot\tau=(\rho c^2-\tau)\dot\Phi-2(p+\tau)/r
\end{equation}
A dot means a derivative with respect to $r$. In the rest of the paper, we choose $8\pi G=1$ and $c=1$. To rewrite the field equations as an autonomous dynamical system for a finite phase space, we define the following variables
\begin{equation}\label{var5}
u=\ln r
\end{equation}
\begin{equation}\label{var2}
P=p r^2
\end{equation}
\begin{equation}\label{var2T}
T=\tau r^2
\end{equation}
\begin{equation}\label{var3}
\mu=\rho r^2
\end{equation}
\begin{equation}\label{var1}
\beta=\tanh (1-b/r)
\end{equation}
\begin{equation}\label{var4}
\theta=\tanh \Phi'
\end{equation}
A prime means a derivative with respect to $u$, an increasing function of $r$. Moreover, the proper radial distance from the throat is
\begin{equation}\label{eqrd}
l=\pm \int^r_{r_0} \frac{dr}{\sqrt{1-b/r}}=\pm \int^r_{r_0} \frac{dr}{\sqrt{\arctanh{\beta}}}
\end{equation}
with $r_0$, the throat size such as $\beta(r_0)=0$. By definition, $l=0$ at the throat, the $+$ sign stands for above it and the $-$ sign, below it. As shown below, $P$ and $T$ can be expressed with $\beta$ and $\theta$ but it is not always possible explicitly for $\mu$ or other free functions. For this reason, we then define 
\begin{equation}\label{var6}
\zeta =\tanh (du/dl)=\pm\tanh (\sqrt{\arctanh\beta}e^{-u})
\end{equation}
$\zeta=0$ at the throat. It is positive or negative just above or below it. But the sign of $\zeta$ does not necessarily stays constant above or below the throat. It changes when it crosses the plane $\zeta=0$ (that is not always at a throat, see \ref{s43}). Obviously the sign of $\zeta$ is the same as this of $dl/dr$. Note that $\beta$ and $\theta$ are two independent variables but not $\beta$ and $\zeta$. Indeed, when we define a trajectory in the phase space by some initial conditions, this on $\zeta$ depends on the initial condition on $\beta$ or vice-versa (see subsection \ref{s12}). $\beta$, $\theta$ and $\zeta$ are three normalised variables in the range from $-1$ to $1$. They allow to rewrite the field equations as an autonomous dynamical system for numerous forms of $\mu$ or other free functions. It comes:
\begin{equation}\label{eq1}
\frac{d\beta}{dl}=\arctanh\zeta\beta'=\arctanh\zeta(1-\beta^2)(1-\arctanh \beta-\mu)
\end{equation}
\begin{eqnarray}\label{eq2a}
\frac{d\theta}{dl}&=\arctanh\zeta\theta'=&\arctanh\zeta\frac{(\theta ^2-1)}{2 \arctanh \beta} \mbox{[}(1-\mu)(\arctanh \theta+1)+\arctanh \beta (-1-\arctanh\theta +\nonumber\\
&&2 \arctanh^2\theta)-2P\mbox{]}\\\nonumber
\end{eqnarray}
\begin{equation}\label{eq4}
\frac{d\zeta}{dl}=\arctanh\zeta\zeta'=-\frac{(1-\zeta^2)\arctanh^2\zeta(-1 + \mu +3 \arctanh\beta)}{2 \arctanh\beta}
\end{equation}
\begin{equation}\label{eq3aa}
-T=-1+\arctanh\beta(1+2\arctanh\theta)
\end{equation}
Equations (\ref{eq1}-\ref{eq4}) form a dynamical system. It is autonomous since $\mu$, $\beta$, $\theta$ and $\zeta$ are some functions of $l$ and it is always possible, at least formally, to write $\mu=\mu(\beta,\theta,\zeta)$. We now assume that the perfect fluid is isotropic, i.e. $p=-\tau$ and define its equation of state $w(r)=p/\rho$. This is the usual case of a cosmological perfect fluid (other choices of perfect fluid are possible with $p\not =-\tau$, see for instance \cite{Kuh17} where $w=\tau/\rho$). Then, the equations (\ref{eq2a}) and (\ref{eq3aa}) rewrite 
\begin{eqnarray}\label{eq2}
\frac{d\theta}{dl}&=&\arctanh\zeta\frac{(\theta ^2-1)}{2 \arctanh \beta} \mbox{[}3+\arctanh \theta -\mu (1+\arctanh \theta)+\nonumber\\
&&\arctanh \beta (-3-5 \arctanh \theta +2 \arctanh^2\theta)\mbox{]}\\\nonumber
\end{eqnarray}
\begin{equation}\label{eq3}
P=-1+\arctanh\beta(1+2\arctanh\theta)
\end{equation}
When $\mu(\zeta)\not = \mu(-\zeta)$, phase space and thus spacetime are different on both sides of the throat. We can then either work with the three equations system $(d\beta/dl,d\theta/dl,d\zeta/dl)$ that allows to plot the trajectories as a function of $l$ in the phase space with $-1<\zeta<1$ or with the three equations system $(\beta',\theta',\zeta')=(d\beta/dl,d\theta/dl,d\zeta/dl)/\arctanh\zeta$ that allows to plot the trajectories as a function of $u$ in two separated phase space, one with $-1<\zeta<0$ and the other with $0<\zeta<1$. In some cases, the dynamical system can be simplified. In particular, when $\mu(\zeta) = \mu(-\zeta)$, phase space trajectories are symmetric with respect to the plane $\zeta=0$. We can then only consider the range $0<\zeta<1$ and the dynamical system $(\beta',\theta',\zeta')$. If moreover $\mu=\mu(\beta,\theta)$, the dynamical system reduces to a set of two equations for $\beta'$ and $\theta'$.\\
Let us give some examples of wormholes models defined by the form of their free function and that can be studied with the above general dynamical system or its simplified forms.
\begin{itemize}
\item If the free function is $b(r)$\cite{Lob05,Cha07}, then from (\ref{eq1A}) we derive $\mu(r)$ and from (\ref{var6}), $\mu(\beta,\zeta)$. If moreover $b(r)$ is invertible to get $r(b)$, then from (\ref{eq1A}), one gets $\mu(r)=\mu(b)=\mu(\beta)$. Then, the dynamical system reduces to a set of two equations for $\beta'$ and $\theta'$.
\item If the free function is $\Phi(r)$\cite{Cat15,Bro03,Kuh06}, we calculate $\Phi'(r)$ and then derive $\theta(\beta,\zeta)$ and $\frac{d\theta}{dl}(\beta,\zeta)$ with (\ref{var6}). We then deduce $\mu(\beta,\zeta)$ from (\ref{eq2}) with $\theta(\beta,\zeta)$ a constraint between the three variables.
\item If the free function is $\rho(r)$\cite{Lob13}, one gets $\mu(\beta,\zeta)$ from (\ref{var6}).
\item If the free function is $p(r)$\cite{Lob13,Rah07}, from (\ref{eq2A}) and (\ref{eq3A}) we derive the Tolman-Oppenheimer-Volkoff equation for static wormhole \cite{Gor08}
$$
\dot p=-\frac{(b+r^3 p) (p+\rho)}{2 r (r-b)}
$$
from which we get $\mu(\beta,\zeta)$ with (\ref{var1}) and (\ref{var6}). If moreover $r(P)$ can be calculated, one derives $\zeta(\beta,\theta)$ from (\ref{eq3}) and thus $\mu(\beta,\theta)$ from $\mu(\beta,\zeta)$.
\item If the free function is $w(r)$, we get $\mu(\beta,\theta,\zeta)=P(\beta,\theta)/w(\zeta,\beta)$ using (\ref{var6}) and (\ref{eq3}).
\item If the free function is $w(z)$ with $z$ the redshift and $w$ is invertible, and if one can calculate $\rho(z)$, then one gets $\rho(w)=\rho(P/\mu)$. Since $r=r(\beta,\zeta)$, then from (\ref{var6}) it is possible to calculate a relation between $\mu=\rho r^2$ and a function of $(\mu,\beta,\theta,\zeta)$. This relation with the equations for $\beta'$, $\theta'$ and $\zeta'$ allow to study the phase space of static wormhole sustained by an equation of state initially given as a redshift function (see subsection \ref{s45}).
\end{itemize}
The above list of model classes that can be studied with the formalism of this paper is not exhaustive. Among others possibilities, one can also derive $\mu(\beta,\theta,\zeta)$ for important barotropic equation of state whose dependence on $r$ is unknown like a constant equation of state (see subsection \ref{s41})\cite{Rah06}, the generalised Chaplygin gas (see subsection \ref{s43})\cite{Ben02}, the Van der Waals equation of state ($p=\frac{8w\rho}{3-\rho}-3\rho^2$ with $w$ a constant)\cite{Kre03}, the quadratic equation of state ($p=\alpha\rho+\beta\rho^2$ with $\alpha$ and $\beta$ some constants)\cite{Ana06}, etc. Some wormhole models are studied in section \ref{s4}.
%------------------------------------------------------------------------------%
\subsection{Relations between some phase space geometrical properties and some wormhole properties}\label{s12}
Some geometrical properties of the phase space $(\beta,\theta,\zeta)$ are related to some physical properties of the wormholes.\\\\
Hence, from the definition of $l$, following a trajectory in the phase space $(\beta,\theta,\zeta)$ from a throat means physically to roll away from this throat. Concerning the plane $\beta=\tanh 1$, it defines the phase space points where the metric becomes spatially flat. A throat in $b=r$ lies on the phase space on the line $(\beta,\zeta)=(0,0)$. Let us show that a throat can only be on this line when $\theta=\pm 1$. Indeed, when $\beta=0$, $d\theta/d\beta$ is diverging, but possibly in $\theta=\pm 1$. Hence the line $(\beta,\zeta)=(0,0)$ cannot be reached by a trajectory but in $\theta=\pm 1$. The points $(\beta,\theta,\zeta)=(0,\pm 1,0)$ are thus the only ones where stand the wormholes throats. From a physical viewpoint, this means that the variation of the gravitational redshift at the throat of a static spherically symmetric wormhole sustained by an isotropic fluid is always diverging (that is not in disagreement with a finite gravitational redshift $\Phi(r_0)$). This generalises the recent result in \cite{Cat16} showing that there is no zero-tidal force static spherically symmetric wormhole ($\theta=0$) sustained by an isotropic perfect fluid. With a non isotropic fluid, things would be different. There are then (among others) zero-tidal force solutions\cite{MorTho88} for which the throat would be in $\theta=0$.\\
Another condition to have a throat at the points $(\beta,\theta,\zeta)=(0,\pm 1,0)$ is given by the flaring-out condition\cite{MorTho88}. It writes for a static wormhole in $r=r_0$
$$
\frac{b-b'}{2b^2}=\frac{e^{-u} \beta'}{2 (\arctanh\beta-1)^2(1-\beta)^2}>0
$$
Hence a phase space trajectory is a wormhole solution if $\beta(r_0)'>0$ at the throat, implying that $\beta>0$ in its neighbourhood. Moreover, we also have $\arctanh^2\zeta/\arctanh\beta = r_0^{-2}$ and thus $d\zeta/dl\not=0$ at the throat. Hence, all the phase space trajectories describing the wormhole solutions pass through a throat at some special points in the phase space but these points are not equilibrium points. Note also that at a throat $d\beta/dl<0$ in $\zeta\rightarrow 0^-$ but $d\beta/dl>0$ in $\zeta\rightarrow 0^+$. These inequalities simply describe a trajectory crossing a throat from $\zeta<0$ to $\zeta>0$ with $\beta\geq 0$. From a physical viewpoint, the flaring-out condition $\beta(r_0)'>0$ also implies from (\ref{eq1}) that $\mu<1$ since at a throat $\beta=0$. Hence we must have $\rho(r_0)<r_0^{-2}$: the larger the throat, the smaller the density to have a wormhole.\\
\\
The curve $P=0$ corresponds to the vanishing of the pressure $p$. It is plotted in the phase space on figure \ref{fig1}. It splits the phase space in several parts $P<0$ and $P>0$, whatever $\zeta$. This curve passes through $(\beta,\theta)=(\tanh 1,0)$ where stands the Minkowski spacetime when $\zeta=0$ (see section \ref{s2}). Its ends for $\beta>0$ come close to $(\beta,\theta)=(1,\tanh(-1/2)^+)$ and $(\beta,\theta)=(0,1)$ and for $\beta<0$ (this part of the space phase shelters the Schwarzschild black hole singularity, see subsection \ref{s40}), close to $(\beta,\theta)=(-1,\tanh(-1/2)^-)$ and $(\beta,\theta)=(0,-1)$. The exact value of $P$ as well as the behaviours of the trajectories at the throats $(\beta,\theta)=(0,\pm 1)$ depend on the value of $\arctanh\beta\arctanh\theta$ in the neighbourhood of the throats where we have the approximations
$$
\arctanh\beta\rightarrow \beta\mbox{ when }\beta\rightarrow 0
$$
$$
\arctanh\theta\rightarrow \mp 1/2\log\epsilon\mbox{ when }\theta\rightarrow \pm 1 \mp \epsilon \mbox{ and }\epsilon\rightarrow 0
$$
$$
\arctanh\zeta\rightarrow \zeta\mbox{ when }\zeta\rightarrow 0
$$
Assuming that $\mu$ does not diverge near the throat where no singularity should stand, it comes 
\begin{equation}\label{th1}
\frac{d\beta}{dl}=\zeta(1-\mu)
\end{equation}
\begin{equation}\label{th2}
\frac{d\epsilon}{dl}=-\zeta\frac{(1-\mu\mp \beta\log\epsilon)\epsilon\log\epsilon}{2\beta}
\end{equation}
\begin{equation}\label{th3}
\frac{d\zeta}{dl}=\zeta^2\frac{(1-\mu)}{2\beta}
\end{equation}
From equations (\ref{th1}) and (\ref{th3}) we recover at the throat that
$$
\zeta\simeq \pm r_0^{-1}\sqrt{\beta}
$$
in agreement with the definition (\ref{var6}) for $\zeta$. If we exclude the possibility of a singularity at the throat, $\beta\log\epsilon$ cannot diverge (see subsection \ref{s31}). Then $\beta\log\epsilon$ tends to a constant $C$ when approaching the throat where, from (\ref{eq3}), $P\rightarrow -1\mp C$. When $C=0$, we derive from (\ref{th1}) and (\ref{th2}) that
$$
\epsilon\simeq e^{const\beta^{-1/2}}
$$
where $const$ is a negative constant (since at the throat $\beta\rightarrow 0^+$) depending on the wormhole model integration constants and parameters. When $C\not= 0$, it comes
$$
\epsilon\simeq e^{C\beta^{-1}}
$$
with $C<0$ such as $\epsilon\rightarrow 0$ when $\beta\rightarrow 0$. These relations allow approximating the behaviour of the trajectories around a throat and can be useful for numerical calculations in its neighbourhood.
\\\\
A last remark is about initial conditions. When solving the original field equations (\ref{eq1A}-\ref{eq3A}), we need three initial conditions corresponding to the three functions $b$, $\Phi$ and $p$. However, when solving the dynamical system for $(\beta,\theta,\zeta)$, we only need two initial conditions for $\beta$ and $\theta$, the initial condition on $\zeta$ being fixed by the one on $\beta$. This is due to the definition of the variable $\theta$. It casts the differential equation (\ref{eq2A}) for $\Phi$ into an algebraic equation (\ref{eq3}) for $P$, $\beta$ and $\theta$. But a third integration constant is recovered if we integrate $\theta$ to get $\Phi$. Moreover, when one chooses some initial conditions $(\beta,\theta,\zeta)=(\beta_1,\theta_1,\zeta_1)$ in $l=l_1\not=0$, $l_1$ must be such as $\beta=0$ (and thus $\zeta=0$) when $l=0$. This fixes the value of the throat size $r_0$ in (\ref{eqrd}). This choice of initial conditions in $l_1\not=0$ corresponds to a unique trajectory in the phase space since there is only one trajectory that contains the point $(\beta_1,\theta_1,\zeta_1)$. However, if one chooses some initial conditions $(\beta_0,\theta_0,\zeta_0)=(0,\pm 1,0)$, i.e. at a throat in $l=0$, there are an infinite number of trajectories that passes through it. Then a unique trajectory is selected by a given value of $r_0$ or any other integration constant.
%------------------------------------------------------------------------------%
\section{Equilibrium points}\label{s2}
In this section, we look for the possible equilibrium points of the dynamical system (\ref{eq1}-\ref{eq4}), i.e. $(d\beta/dl,d\theta/dl,d\zeta/dl)$. For the reader who wants to avoid this mathematical section, its results are summarised on figure \ref{fig1}. We do not determine the equilibrium points stability that can only be found once the dynamical system is fully specified. However, some general remarks can be made about source and sink points. Hence, a trajectory has not always a source or a sink point. Some of them can begin or end at a non equilibrium point, like some singularities (see an example in subsection \ref{s40}). Note also that the physical meaning of a source or a sink point for a static wormhole trajectory is different from the one of an isotropic and homogeneous cosmology. When in cosmology some phase space trajectories depend on the proper time parameter $t$, a source or a sink point then takes place respectively at early or late times. Physically, an observer can thus only follow a trajectory from this source to this sink to respect the causality. For a static wormhole, the parameters $l$ and $r$ of the trajectories, are related to the distance to a throat and its axis. An observer can thus physically go to a sink as well as to a source point on a wormhole trajectory.\\
In the following, the three sets of equilibrium points for the equations (\ref{eq1}-\ref{eq4}) correspond to the three possibilities such as $d\beta/dl=0$ in (\ref{eq1}), respectively $\beta=\tanh(1-\mu)$, $\beta=\pm 1$ or $\zeta=0$. Note that the first equality is an equation $\beta=\tanh(1-\mu(\beta,\theta,\zeta))$ that can correspond to several values of $\beta$.
%------------------------------------------------------------------------------%
\subsection{Set 1: $\beta=\tanh(1-\mu)$}\label{s21}
\begin{figure}[ht]
\centering
\includegraphics[width=8cm]{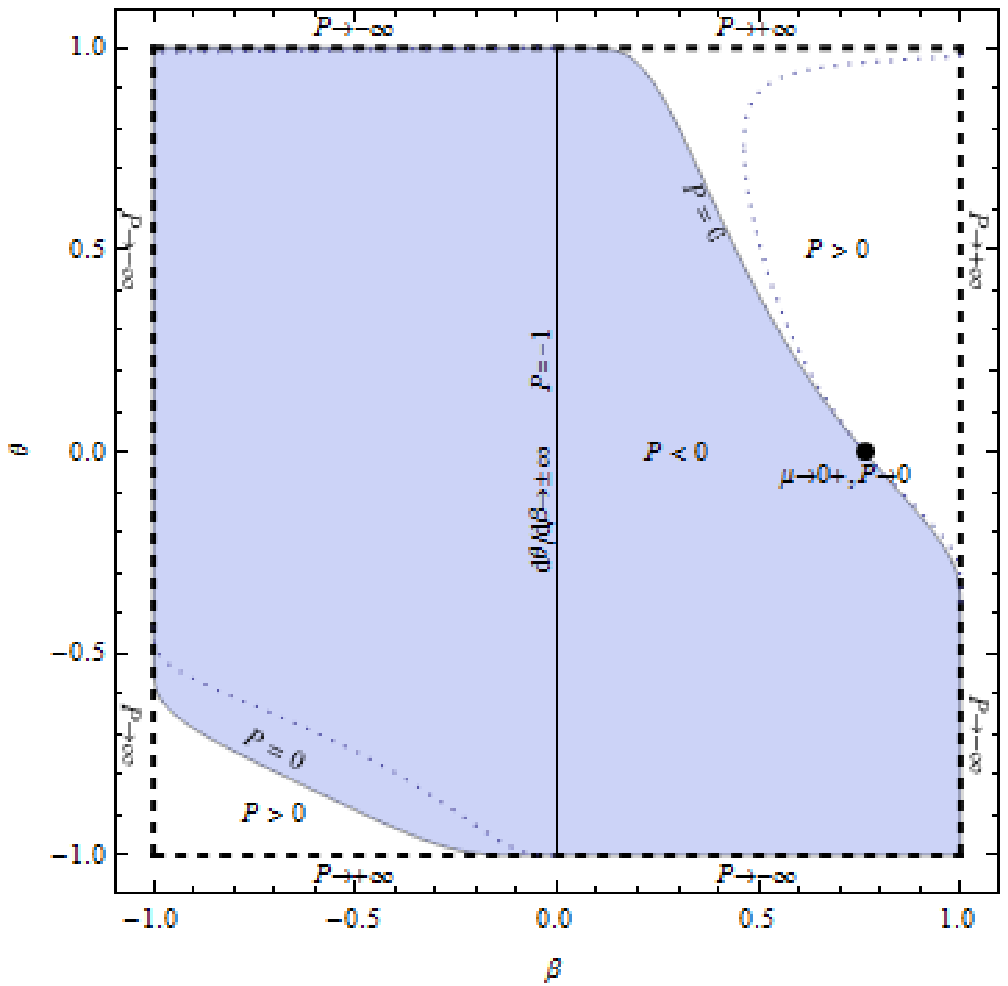}
\includegraphics[width=8cm]{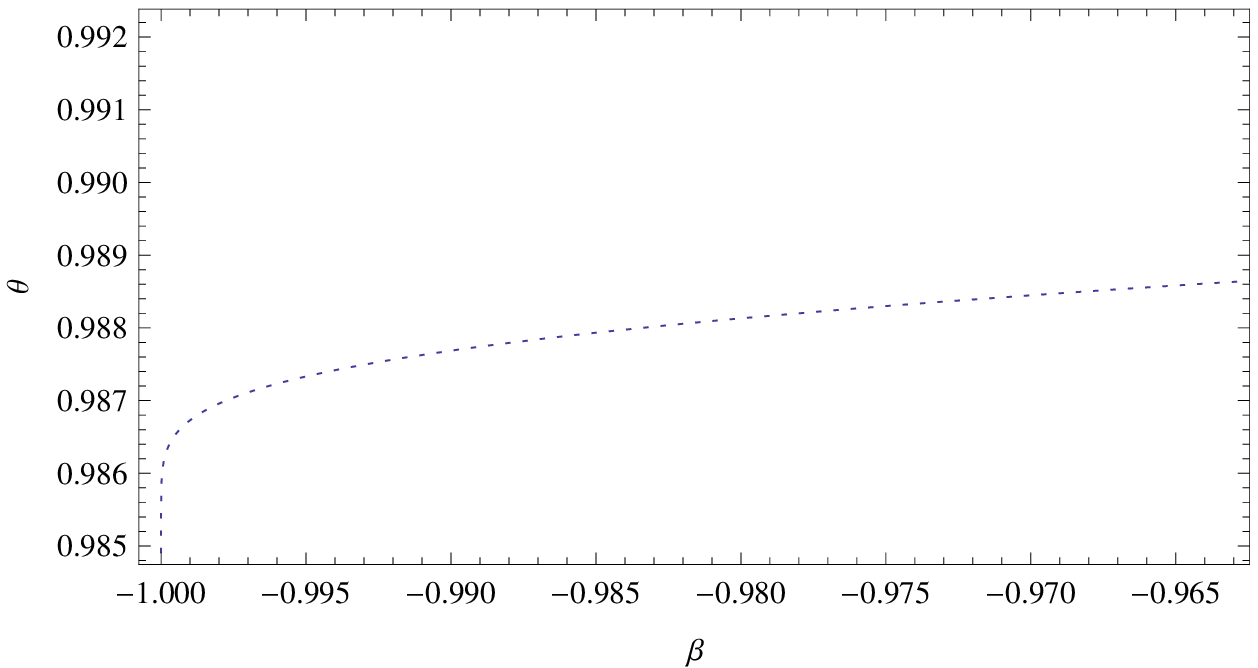}
\caption{\scriptsize{\label{fig1} Some properties of the phase space $(\beta,\theta,\zeta)$ projected on the surface $(\beta,\theta)$ for a static wormhole sustained by a perfect fluid in General Relativity. The dark (white) area is such that $P<0$ (respectively $P>0$). The dotted curves and dashed lines contain the phase space points that can be some equilibrium points (on the planes $\zeta=0$ or $\pm 1$ if $\mu$ depends on $\zeta$). The second figure shows these possible equilibrium points near $\theta=1$ when $\beta<0$. The large point in $(\beta,\theta)=(\tanh 1,0)$ stands for the Minkowski spacetime. Singularities can be found wherever $\mu$ diverges and, independently on $\mu$, on the dashed lines $\theta=\pm 1$ and $\beta=\pm 1$. On the vertical line $\beta=0$, the slope $d\beta/d\theta$ diverges but not necessarily in $\theta=\pm 1$ where stand throats, singularities or the Schwarzschild event horizon.}}
\end{figure}

In this subsection, we consider that $d\beta/dl=0$ when $\beta=\tanh(1-\mu)$. Then from (\ref{eq4}), $d\zeta/dl=0$ implies that the equilibrium points are on the plane $\zeta=0$ (then, from the definition (\ref{var6}) for $\zeta$, $\beta=0$ with $\mu=1$ or/and $r$ is diverging) or $\zeta=\pm 1$ (then $\beta=\pm 1$ with $\mu=0$ or/and $r=0$. We have a singularity (see section \ref{s31})).
Last, from (\ref{eq2}), $d\theta/dl=0$ when $\zeta=0$ but also if
$$
\theta=\pm 1
$$
or
\begin{equation}\label{eqP}
\theta=\tanh(1\pm\frac{\sqrt{2\mu^2-3\mu+1}}{\mu-1})
\end{equation}
The first subset with $\theta=\pm 1$ can correspond to some singularities (see section \ref{s3}). The second subset (\ref{eqP}) of equilibrium points is real when $\mu\not\in\left[1/2,1\right]$. It defines three dotted curves in the plane $(\beta,\theta)$ (shown on figure \ref{fig1}) where these points can be located. More specifically, when the density tends to vanish ($\mu\rightarrow 0$) or equivalently $\beta\rightarrow \tanh 1$, spacetime becomes spatially flat and we get the following values for $\theta$:
\begin{itemize}
\item $\theta=0$. This is the Minkowski solution when $\Phi$ tends asymptotically ($r\rightarrow +\infty$) to a constant and $\zeta$ to $0$. Since in $(\beta,\theta)=(\tanh 1,0)$, $\mu$ and $P$ are vanishing, this point is reached if the density $p$ and the pressure $\rho$ vanish faster than $r^{-2}$ as indicated by the definitions (\ref{var2}) and (\ref{var3}) for $P$ and $\mu$. See subsection \ref{s42} for a wormhole model tending asymptotically to the Minkowski spacetime. Let us remark that the Minkowski solution is reached when $d\Phi/du\rightarrow 0$ and not simply $d\Phi/dr\rightarrow 0$. This is because $d\Phi/dr$ have to decrease faster than $1/r$ such as $\Phi(r)$ tends to a constant when $r$ is diverging. This implies $r d\Phi/dr =d\Phi/du \to 0$.
\item $\theta=\tanh 2$. This point corresponds to a spatially flat Universe different from the Minkowski one. Especially, $\Phi'\rightarrow 2$ implies that the gravitational redshift $\Phi$ diverges as $2\ln r$ far from the throat or equivalently, the metric function $e^{2\Phi}$ as $r^4$.
\item $\theta=\pm 1$. As shown on section \ref{s3}, these points can correspond to singularities when they are reached at finite $u$. Such a singularity in a spatially flat spacetime is not a curvature singularity (see subsection \ref{s31}) produced by a strong gravity field. But note that at these points the variation of the gravitational redshift diverges.
\end{itemize}
%------------------------------------------------------------------------------%
\subsection{Set 2: $\beta=\pm 1$}\label{s22}
In this subsection, we consider that $d\beta/dl=0$ when $\beta=\pm 1$. As shown in section \ref{s31}, this set can correspond to some singularities when $\beta=\pm 1$ is reached at finite $u$. Following the definition (\ref{var6}) for $\zeta$, $d\zeta/dl=0$ implies $\zeta=0$ (with then $r$ diverging) or $\zeta=\pm 1$. Moreover, $d\theta/dl=0$ when $\zeta=0$ but also when
$$
\theta=\pm 1
$$
or
$$
\theta=\tanh\frac{\mu-1+5 \arctanh \pm 1+\sqrt{(\mu-1)^2+2 (9 \mu -17) \arctanh \pm 1+49 \arctanh^2 1}}{4 \arctanh \pm 1}
$$
or
$$
\theta=\tanh\frac{\mu-1+5 \arctanh \pm 1-\sqrt{(\mu-1)^2+2 (9 \mu -17) \arctanh \pm 1+49 \arctanh^2 1}}{4 \arctanh \pm 1}
$$
In the above expressions, we leave the diverging value $\arctanh \pm 1$ since to evaluate the corresponding limit for $\theta$, the exact form of $\mu$ is necessary. In the particular case $\mu<<\arctanh 1$, we get the equilibrium points
\begin{equation}\label{eqP2A}
(\beta,\theta)=(\pm 1,-\tanh\frac{1}{2})
\end{equation}
\begin{equation}\label{eqP2B}
(\beta,\theta)=(\pm 1,\tanh 3)
\end{equation}
%------------------------------------------------------------------------------%
\subsection{Set 3: $\zeta=0$}
In general, $\zeta=0$ implies $d\beta/dl=0$, $d\theta/dl=0$ and $d\zeta/dl=0$ but at the points $(\beta,\theta,\zeta)=(0,\pm 1,0)$ (that are not necessarily some throats when $\beta'<0$) since then $\arctanh\zeta/\sqrt{\arctanh\beta}=r_0^{-1}$ (see the definition (\ref{var6}) of $\zeta$) and thus $d\zeta/dl\not = 0$. There is also no equilibrium  when $\mu$ is diverging but there is then a singularity. Consequently, no wormhole trajectory can cross the plane $\zeta=0$ but in $(\beta,\theta,\zeta)=(0,\pm 1,0)$.\\\\
The positions of all the possible equilibrium points projected on the surface $(\beta,\theta)$ are shown on figure \ref{fig1}. They all stand on the planes $\zeta=0$ or $\pm 1$.
%------------------------------------------------------------------------------%
\section{Singularity}\label{s3}
In this section, we determine the location of (curvature) singularities in the phase space and the geodesic equation for spacetimes described by these trajectories. For readers who want to jump this section, its results are summarised on figure \ref{fig1}.
%------------------------------------------------------------------------------%
\subsection{Singularity}\label{s31}
We are interested by curvature singularities, i.e. Ricci and Weyl singularities. To find them, we consider the Ricci scalar
$$
R=g_{\mu\nu}R^{\mu\nu}=e^{-2u}\left[3+\mu-3\arctanh\beta(1+2\arctanh\theta)\right]
$$
but also
\begin{eqnarray}
R_{\gamma \delta }R^{\gamma \delta }&=&e^{-4 u} \mbox{[}3+\mu ^2-6 \arctanh\beta (1+2 \arctanh\theta)+\nonumber\\
&&3 \arctanh^2\beta (1+2 \arctanh\theta)^2\mbox{]}\nonumber
\end{eqnarray}
and the Kretschman scalar
\begin{eqnarray}
R_{\gamma \delta \nu \xi }R^{\gamma \delta \nu \xi }&=&e^{-4 u} (15+\mu  (3 \mu-10 )+2 (\mu-3)\arctanh\beta (5+2 \arctanh\theta)+ \nonumber\\
&&3 \arctanh^2\beta \left[5+4 \arctanh\theta(1+\arctanh\theta)\right])\nonumber
\end{eqnarray}
When $\mu$ diverges at finite $u$, we have a Ricci singularity. But the scalars can also diverge, whatever $\mu$, when $\beta=\pm 1$ and/or $\theta=\pm 1$ (these values resulting from the compactification) but in $(\beta,\theta)=(0,\pm 1)$ if we have a throat. If this occurs at finite $u$ and $\mu$, we have a Weyl singularity for which the tidal forces become infinitely large\cite{Ugg06}. Note that even if a special form of $\mu$ is chosen to cancel the divergence of $R$, it will not avoid this of the other scalars and vice-versa.\\
\\
Since the $(t, r,\theta,\phi)$ coordinates of the metric (\ref{met}) we use in this paper do not cover any horizon, we do not look for their possible positions in the phase space. However, in section \ref{s40}, we show how to extend this work to the representation of black holes in the phase space by examining the Schwarzschild black hole. Then, we can locate its event horizon in the phase space since it has been shown in other coordinates (Kruskal-Szekeres for instance) that it stands in $\Phi\rightarrow -\infty$. In a future work, we hope to describe black holes representation in a phase space where the notion of horizon could be appropriately treated.
%------------------------------------------------------------------------------%
\subsection{Geodesic equation}\label{s33}
To confirm the presence of a singularity when some scalars diverge, we want to check when a spacetime is geodesically incomplete. We thus calculate the geodesic equation giving $u$ as a function of the proper time $\tau$ for a test particle in a spacetime described by a phase space trajectory. Since we are interested by static spherically symmetric wormhole, we only consider the equatorial plane $\theta=\pi/2$ (the $\theta$ of the spacetime metric, not the variable of the dynamical system). From the geodesic equations, we get the conserved quantities\cite{Cat17}
$$
\frac{dt}{d\tau}=\frac{E}{m} e^{-2\Phi}
$$
$$
\frac{d\phi}{d\tau}=A e^{-2u}
$$
with $E$ and $A$, respectively the conserved energy and angular momentum of the test particle with mass $m$. From the relation for four-velocity $U$ when considering time-like geodesics, we have $U^\mu U_\mu=-1$ from which we derive
$$
A^2 e^{-2u}+\frac{E^2}{m^2} e^{-4\Phi(u)}\left[-e^{2\Phi(u)}+\frac{e^{2u}(du/d\tau)^2}{\arctanh\beta(u)}\right]=-1
$$
For sake of simplicity, we now choose $E=A=1$ (and thus non radial geodesics since $A\not =0$). Then, from this last equation, we deduce
\begin{equation}\label{geo}
e^{2\Phi(u)}=\frac{2 e^{2 u} (du/d\tau)^2}{\arctanh\beta(u)\pm\sqrt{\arctanh\beta(u) \left[\arctanh\beta(u)-4 (1+e^{2 u}) (du/d\tau)^2\right)]}}
\end{equation}
Equation (\ref{geo}) contains two solutions for $u(\tau)$, depending on the $\pm$ sign. This is due to the terms $e^{-2\Phi}$ and $e^{-4\Phi}$ in the previous equation. The solution with the plus (minus) sign describes trajectories with $\beta>0$ (respectively $\beta<0$) and allows to calculate $u(\tau)$ for phase space trajectories.\\
From a technical viewpoint, to calculate $u(\tau)$, we first get a (generally) numerical solution for $(\beta(u),\theta(u))$ from the dynamical system (\ref{eq1}-\ref{eq4}). Since $\theta$ depends on $\Phi'$, we differentiate (\ref{geo}), introduce the solution for $(\beta(u),\theta(u))$ in this equation for $\Phi'$, choose some initial conditions for $u(\tau)$ and then solve the differential equation for $u(\tau)$. This method is applied in the next section when we need to confirm the presence of some singularities.
%------------------------------------------------------------------------------%
\section{Some wormholes phase spaces}\label{s4}
In this section, we present the phase space of several wormhole models:
\begin{itemize}
\item a vanishing density
\item a generalised Chaplygin gas
\item a constant equation of state
\item a linear equation of state
\item a Chevallier-Polarski-Linder equation of state
\item a Navarro-Frenk-White profile density
\item a toy model with an asymptotically vanishing pressure
\end{itemize}
From a mathematical viewpoint, we show how to apply the dynamical system formalism to get and classify the solutions of the above models. All these solutions are summarised in table \ref{tab2}. From a physical viewpoint, the vanishing density model (see subsection \ref{s40}) shows how to possibly extend this work to consider static black hole trajectories. Dark energy and matter models (see subsection \ref{s4a}) show that usual models of dark fluids cannot form wormholes naturally since they cannot be asymptotically Minkowski flat. Finally, a toy model with an asymptotically vanishing pressure (see subsection \ref{s42}) shows that a static wormhole can be flat on both sides of its throat contrary to some recent results\cite{Bro17}. 
\begin{table}
\begin{center}
\begin{tabular}{|c|c|c|c|}
\hline
Model & Parameter range & Throat & End point \\
\hline
$\mu=0$ &  & $(0,1)$ & $\theta=-1$ with $P<0$\\
if $P<0$& 	& 			& 		singularity		\\
\hline
Chaplygin & $(A,\alpha)>0$ & $(0,1,0)$ & $(0,-1,0)$ with $P<0$\\
if $A<r_0^{-2(1+\alpha)}$	& 	& 			& 	Cyclic								\\
\hline
Constant eos & $w<-3-2\sqrt{2}$& $(0,1)$ & $(0,-1)$ with $P<0$, $r\rightarrow \infty$ \\
if $w\not\in\left[-1,0\right]$		&					& $(0,1)$ & $(\tanh\frac{(1+w)^2}{w^2+6w+1},\tanh\frac{2w}{1+w})$ with $P>0$, $r\rightarrow \infty$ \\
			& $-3-2\sqrt{2}<w<-1$& $(0,1)$ & $(0,-1)$ with $P<0$, $r\rightarrow \infty$ \\
			& 					& $(0,1)$ & $(1,\tanh\frac{1+3w}{w-1})$ with $P>0$, $r\rightarrow \infty$ \\
			& $0<w<1$& $(0,1)$ & $(1,\tanh\frac{1+3w}{w-1})$ with $P<0$, $r\rightarrow \infty$ \\
			& 			& $(0,-1)$ & $(1,\tanh\frac{1+3w}{w-1})$ with $P<0$, $r\rightarrow \infty$ \\
			& $1<w$& $(0,1)$ & $(1,-1)$ with $P<0$, $r\rightarrow \infty$ \\
\hline
Linear eos & $-1.5<w_0<-1,-0.5<w_1<0$ & $(0,1,0)$ & $(0,-1,0)$ with $P<0$, $r\rightarrow \infty$\\
NA	& 			& 			& 						\\
\hline
CPL eos& $-1.5<w_0<-1,-0.5<w_1<0$ & $(0,1,0)$ & $(0,-1,0)$ with $P<0$, $r\rightarrow \infty$\\
NA	& 			& 			& 						\\
\hline
NFW & $(\rho_r,r_s)>0$ & $(0,1,0)$ & $(\tanh 1,1,0)$ with $P>0$, $r\rightarrow \infty$, flat\\
if $1-\frac{\pm r_0r_s^3\rho_s}{\pm r_0+r_s}^2>0$	& 					& $(0,1,0)$ & $(\tanh 1,-1,0)$ with $P<0$, $r\rightarrow \infty$, flat\\
\hline
$p=p_0r^{-n}$ & $n>2$ & $(0,1)$ & $\theta=0$ with $P<0$\\
	if $p_0r_0^{2-n}>-1$& 		& 			& 		singularity		\\
			& 		& $(0,1)$ & $(\tanh 1,0)$ with $P>0$, $r\rightarrow \infty$\\
			& 		& 			& 		Minkowski		\\
\hline
\end{tabular}
\caption{The first column gives the model and conditions to have a wormhole. The second column gives the range of parameters we have considered. The third column gives the location of the throats in the 2 or 3 dimensional space phase $(\beta,\theta,\zeta)$. The fourth column gives the behaviour far from the throat of the trajectories. None of the usual dark energy and dark matter models produces an asymptotically flat Minkowski wormhole.}
\label{tab2}
\end{center}
\end{table}
%------------------------------------------------------------------------------%
\subsection{Vanishing density: a wormhole exact solution and a Schwarzschild black hole trajectory}\label{s40}
\begin{figure}[h]
\centering
\includegraphics[width=6cm]{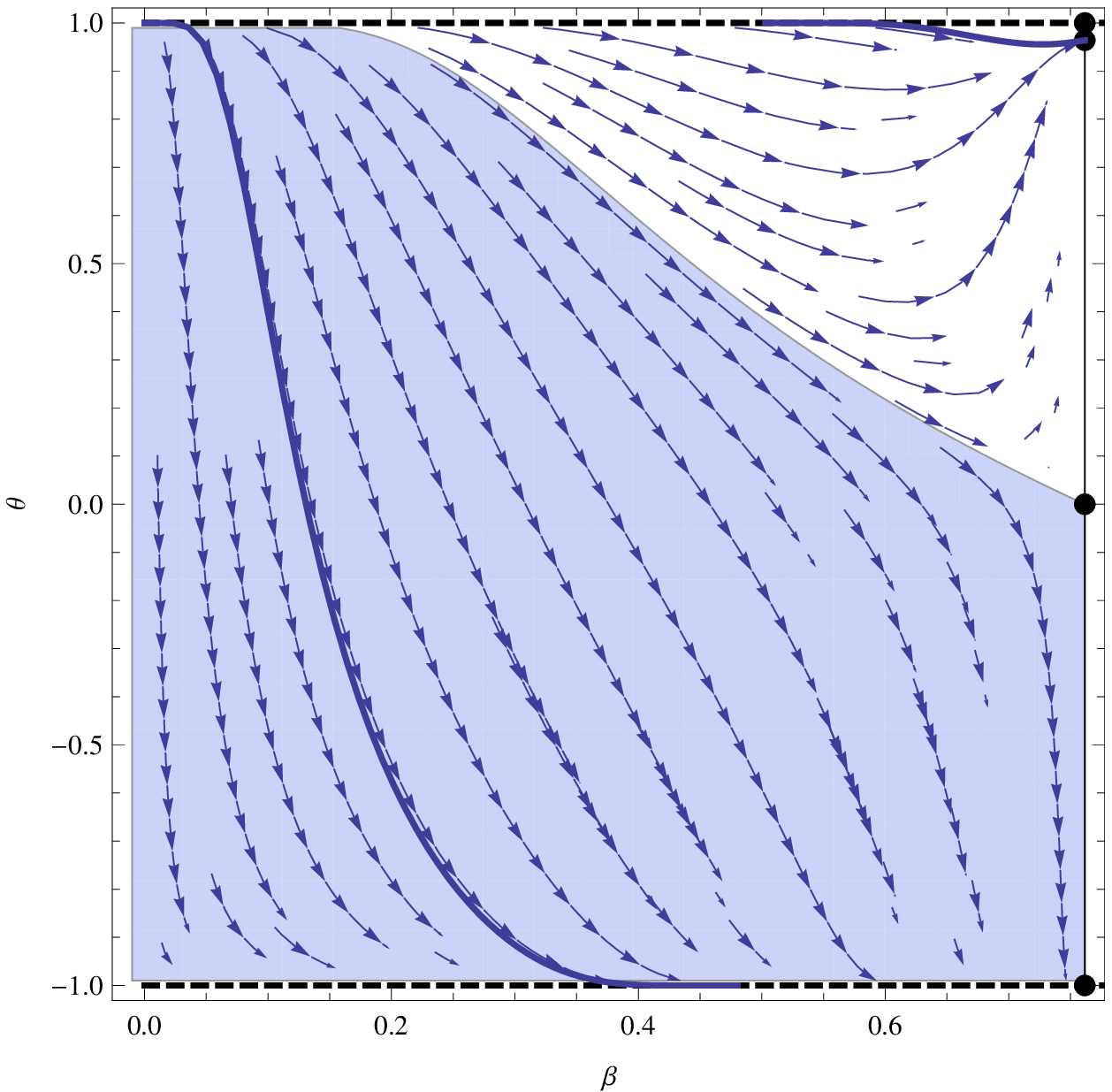}
\includegraphics[width=6cm]{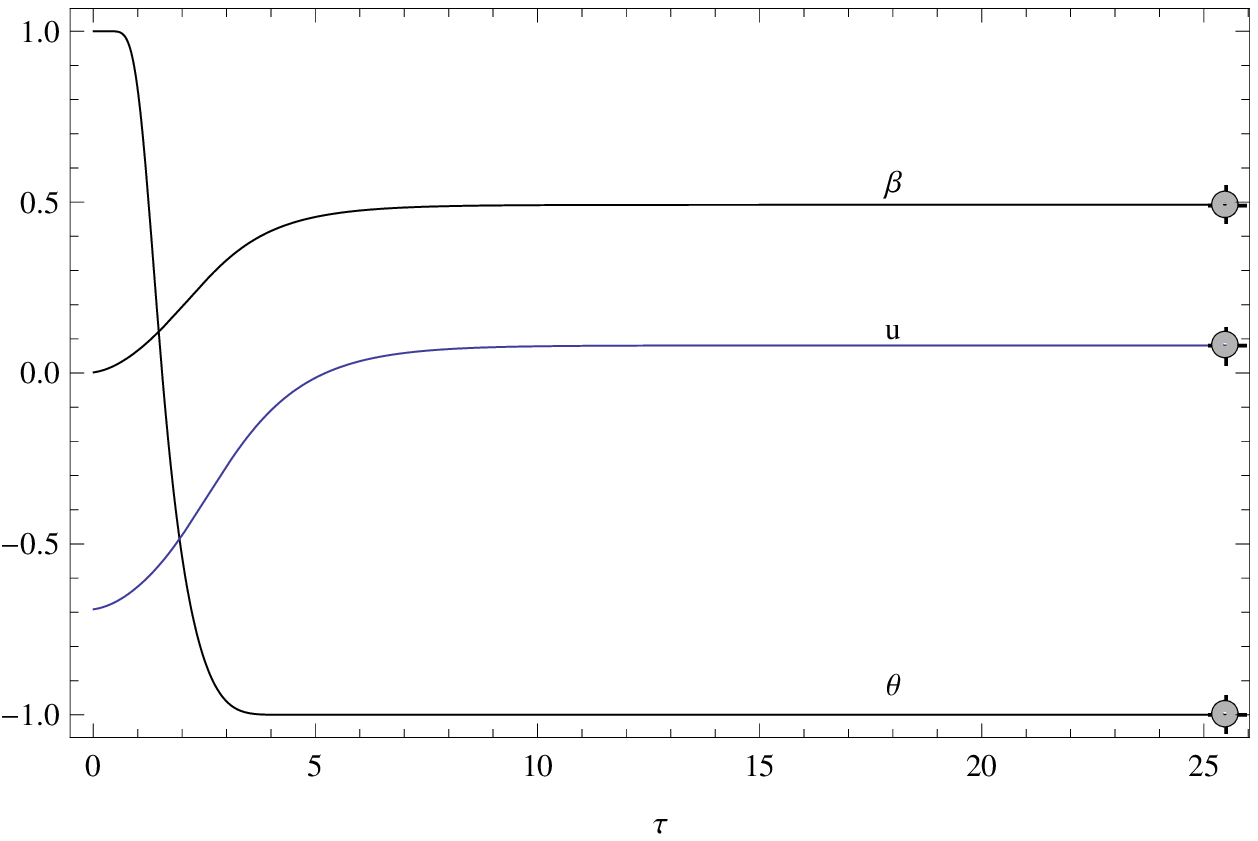}
\caption{\scriptsize{\label{fig2}First graph: Phase space of a wormhole with vanishing density. The equilibrium points are in bold on the line $\beta=\tanh 1$. The thick trajectory is defined by $r_0=0.5$, $p_1=1$ and $\Phi_1=1$. There are two different parts. The first one describes a wormhole: the throat is in $(\beta,\theta)=(0,1)$ and it ends at a finite $r$ and $\tau$ in $(\beta,\theta)=(\beta_1^-,-1)$ with a pressure that tends to $-\infty$. The second one, that does not describe a wormhole, starts in $(\beta,\theta)=(\beta_1^+,+1)$ where the pressure tends to $+\infty$ and then goes to $(\beta,\theta)=(\tanh 1,\tanh 2)$ at infinite $r$. Second graph: $u(\tau)$, $\beta(\tau)$ and $\theta(\tau)$ for the left part of the thick trajectory representing a wormhole on the first graph. The gray points indicate the ends of the curves $u(\tau)$ , $\beta(\tau)$ and $\theta(\tau)$ in $\tau=25.5$ and $u=0.08$.}}
\end{figure}

\begin{figure}[h]
\centering
\includegraphics[width=8cm]{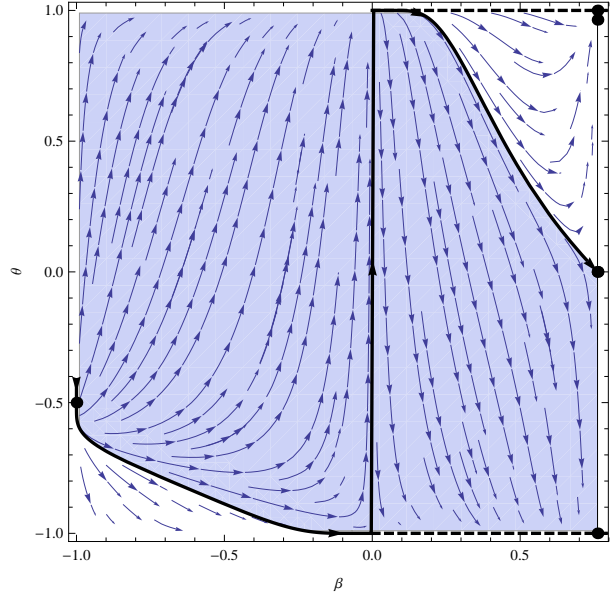}
\caption{\scriptsize{\label{fig8}phase space for $\mu=0$ but with $-1<\beta<1$. The Schwarzschild black hole trajectory is plotted as a bold trajectory. This is the only one reaching asymptotically the Minkowski equilibrium point. The trajectory is discontinued at the event horizon, in $(\beta,\theta)=(0,\pm 1)$.}}
\end{figure}

The first model we choose to study is defined by a vanishing density $\mu=0$. From a mathematical viewpoint, it can be solved exactly with the original equations system (\ref{eq1A}-\ref{eq3A}). This allows to check its consistency with the dynamical system (\ref{eq1}-\ref{eq4}). From a physical viewpoint, the $\mu=0$ model contains as a particular case ($P=0$) the Schwarzschild black hole. This gives us the opportunity to examine how a phase space describing a static spherically symmetric wormhole could also be used to study a static spherically symmetric black hole for future work and to better understand the relations between these two types of objects.\\\\
The exact solution of the equations system (\ref{eq1A}-\ref{eq3A}) is:
$$
b=r_0
$$ 
$$
p=p_1 e^{-\Phi}
$$
\begin{eqnarray}
e^{-\Phi}&=&-\frac{1}{2 c^4 \sqrt{r} \sqrt{-r_0+r}}(-30 r_0^2 G \pi \sqrt{r} \sqrt{-r_0+r} p_1+4 G \pi r^{5/2} \sqrt{-r_0+r} p_1\nonumber\\
&&-r \Phi_1+r_0(10 G \pi r^{3/2} \sqrt{-r_0+r} p_1+\Phi_1)-30 r_0^2 G \pi (r_0-r) p_1\times\nonumber\\
&&\log\left[\sqrt{r}+\sqrt{-r_0+r}\right])
\end{eqnarray}
with $r_0$ the throat size, $p_1$ and $\Phi_1$, three integration constants. At the throat, we have $e^{-\Phi}\rightarrow -(r_0^2 p_1)^{-1}$. The special case $p_1=0$ corresponds to the Schwarzschild black hole (see the end of this subsection) with an event horizon in $\Phi\rightarrow -\infty$. Let us first consider that $p_1\not =0$. A solution defined by $r_0=0.5$, $p_1=1$ and $\Phi_1=1$ is plotted in the phase space on the first graph of figure \ref{fig2}. The comparison between this solution and the phase space trajectories shows the consistency of the dynamical system approach with respect to the above exact solution calculated with equations (\ref{eq1A}-\ref{eq3A}).\\
The phase space trajectories representing wormhole solutions are such as $\beta<\tanh 1$ since on the line $\beta=\tanh 1$, $d\theta/d\beta$ diverges. Hence, trajectories coming from the throat cannot cross this line. The dynamical system has four equilibrium points, all on the line $\beta=\tanh 1$. They thus correspond to spatially flat solutions. They are
\begin{itemize}
\item $(\beta,\theta)=(\tanh 1,-1)$ that is a sink.
\item $(\beta,\theta)=(\tanh 1,1)$ that is a saddle.
\item $(\beta,\theta)=(\tanh 1,0)$ that is a saddle Minkowski point.
\item $(\beta,\theta)=(\tanh 1,\tanh 2)$ that is a sink.
\end{itemize}
At the throat in $(\beta,\theta)=(0,1)$, $\beta'\rightarrow 1>0$ in agreement with the flaring-out condition, and the absence of equilibrium point there. We also have that $P=-1$ and thus the equation of state $p/\rho\rightarrow -\infty$, in agreement with violation of the weak energy condition to have a wormhole.\\
A wormhole trajectory behaves in the following way. It starts at a throat in $(\beta,\theta)=(0,1)$. Further from the throat, at a finite coordinate $r=r_1>r_0$, the trajectory reaches the line $\theta\rightarrow -1$ in $r_1^-$. The pressure and the Ricci scalar $R$ diverge. The geodesic equation (\ref{geo}) shows that the coordinate $r$ ends in $r_1^-$ at a finite time $\tau$. There is thus a singularity at finite distance and time from the wormhole throat. A geodesic solution $u(\tau)$ for the special case $r_0=0.5$, $p_1=1$ and $\Phi_1=1$ is plotted on the second graph of figure \ref{fig2} with a singularity occurring in $\tau=25.5$ and $u=0.08$. In $r_1^+$, the trajectory restarts with the same $\beta$ as in $r_1^-$ but in $\theta\rightarrow +1$. At this point, anew, the pressure and $R$ are diverging. We thus also have a singularity in $r_1^+$ that is confirmed, as for $r_1^-$, by the geodesic equation (\ref{geo}). For larger $r$, the trajectory goes to the saddle points and then to the sink point $(\beta,\theta)=(\tanh 1,\tanh 2)$ with $r\rightarrow \infty$, where the metric becomes spatially flat. This second part of the trajectory with $r>r_1^+$ does not belong to a wormhole solution since it is disconnected from the trajectory with $r_0<r<r_1^-$. Physically the $\mu=0$ model thus generate a wormhole that relates two singularities.\\
Although this paper is dedicated to wormholes, let us have a look now on the representation of the Schwarzschild black hole defined by $\mu=P=0$ (i.e. $p_1=0$) in such a phase space. Its singularity is in $r=0$, i.e. $\beta=-1$. As is well known, its event horizon occurs in $\Phi\rightarrow -\infty$ at finite $u$, that means $\Phi'\rightarrow \pm\infty$ or $\theta\rightarrow \pm 1$. But we know that on these lines we also have a singularity but possibly in $\beta=0$. Hence, the event horizon of the Schwarzschild black hole can only take place in $(\beta,\theta,\zeta)=(0,\pm 1,0)$. It follows that we have to consider the whole phase space with $-1<\beta<1$ to plot the trajectory of the Schwarzschild black hole (see \cite{Dea99} for a dynamical system analysis of Schwarzschild orbital dynamics).  It is represented in bold on figure \ref{fig8} and can be described as follows. It starts at a singularity in $(\beta,\theta)=(-1,-\tanh 1/2)$, which is a source equilibrium point defined in (\ref{eqP2A}), and simply follows the curve $P=0$. The trajectory is discontinued (as the curve $P=0$) when crossing the event horizon in $(\beta,\theta)=(0,\pm 1)$. This reflects the usual discontinuity in $r=2M$ (with $M$ the black hole mass) of the Schwarzschild metric in $(r,t)$ coordinates. The trajectory then continues until reaching the Minkowski equilibrium point. Others trajectories with $P\not =0$, starting at the same singularity, are not black holes. They stay in the area $\beta<0$ and end in another singularity in $\theta=\pm 1$ at a constant value of $\beta<0$ (in a similar way to the wormhole trajectories with $\beta>0$ and $\theta=-1$). Hence, the only type of black hole that can exist with a fluid having a vanishing density is the Schwarzschild black hole. We will not go further in this paper about considerations on black holes but it could be interesting to use this phase space to determine the possible formation of black holes with various dark fluids.
%------------------------------------------------------------------------------%
\subsection{Usual cosmological dark fluids: no asymptotically Minkowski flat wormhole}\label{s4a}
In this subsection, we analyse some static spherically symmetric wormholes sustained by models of dark fluids usually used to describe the effects of dark energy and dark matter on Universe. These models are:
\begin{itemize}
\item a generalised Chaplygin gas (dark energy, three dimensional phase space)
\item a constant equation of state (dark energy, two dimensional phase space)
\item a linear equation of state (dark energy, usually given as a redshift function, three dimensional phase space)
\item a Chevallier-Polarski-Linder equation of state (dark energy, usually given as a redshift function, three dimensional phase space)
\item a Navarro-Frenk-White profile density (dark matter, three dimensional phase space)
\end{itemize}
From a mathematical viewpoint, these models illustrate how to apply the dynamical system formalism of this paper to large classes of wormholes, even when their equation of state is known as a redshift function. From a physical viewpoint, we classify the wormhole solutions of these dark fluids and show that none of them is asymptotically Minkowski flat. These isotropic dark fluids should thus not form static wormholes naturally.
%------------------------------------------------------------------------------%
\subsubsection{Generalised Chaplygin gas}\label{s43}
\begin{figure}[t]
\centering
\includegraphics[width=8cm]{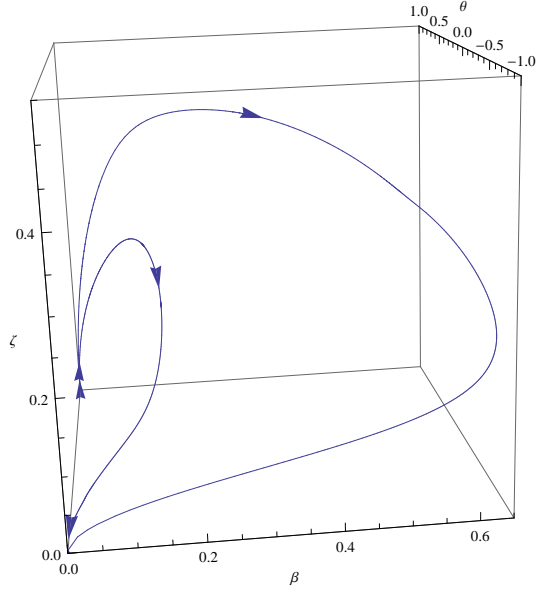}
\caption{\scriptsize{\label{fig15}Some wormhole trajectories in the $(\beta,\theta,\zeta)$ phase space, sustained by a Chaplygin gas with $A=0.1$ and $\alpha=1$. Each curve ends in $(\beta,\theta,\zeta)=(0,-1,0)$ at finite $l$ and has a symmetrical part with respect to the $\zeta=0$ plane. We get similar figure with other values of $A$ and $\alpha$.}}
\end{figure}

The pressure of a generalised Chaplygin gas\cite{Ben02} is given by
$$
p=-\frac{A}{\rho^\alpha}
$$
with $A$ and $\alpha$ two positive constants. The pressure is thus negative. We deduce that
$$
\mu=(\frac{\arctanh\beta}{\arctanh^2\zeta})^\frac{1+\alpha}{\alpha}(\frac{A}{1-\arctanh{\beta}(1+2\arctanh{\theta})})^{\frac{1}{\alpha}}
$$
The density is thus positive. Approximating $\arctanh\beta$ and $\arctanh\theta$ by respectively $\beta$ and $\mp 1/2\log\epsilon$ near the throat (see subsection \ref{s12}), we derive that $\beta'>0$ implies $A<r_0^{-2(1+\alpha)}$ in agreement with \cite{Lob06A}. Since $\mu$ depends on $\zeta$, we have to use the three dimensional dynamical system for $(\beta,\theta,\zeta)$ to study the phase space. However, since $\mu(\zeta)=\mu(-\zeta)$, the phase space for both sides of the throat is symmetric with respect to the plane $\zeta=0$ and we thus only consider the range $0<\zeta<1$. Moreover, when $\beta=\tanh 1$, $d\beta/dl<0$ and thus $\beta$ is decreasing for this special value. This implies that the wormhole trajectories coming from the throat in $\beta=0$ cannot cross the plan $\beta=\tanh 1$. Numerical simulations show then that the throat is in $(\beta,\theta,\zeta)=(0,1,0)$ and the trajectories go to $(\beta,\theta,\zeta)=(0,-1,0)$ at finite $l$, density and curvature as shown on figure \ref{fig15}. By symmetry with respect to the $\zeta=0$ plane, we thus get a closed trajectory. This describes a succession of identical wormholes connected by their throats, i.e. a cyclic structure repeating infinitely.
%------------------------------------------------------------------------------%
\subsubsection{Constant equation of state}\label{s41}
A constant equation of state $w=p/\rho=P/\mu$ has observational preferred value around $w=-1$, i.e. near the value of a cosmological constant (see for instance \cite{Sta17}). Wormholes with similar equation of state have been studied in \cite{Rah06,Cat15}, with $p\not =-\tau$ in \cite{Sus05,Lob06,Kuh16} or with $\tau=w\rho$ in \cite{Kuh17}. From the pressure (\ref{eq3}), we derive that a constant equation of state $w$ is obtained when
$$
\mu=\left[-1+\arctanh\beta(1+2\arctanh\theta)\right]/w
$$
$\mu$ does not depend on $\zeta$ and the dynamical system thus reduces to the differential equations for $\beta'$ and $\theta'$. Wormhole throats for this model are in $(\beta,\theta)=(0,\pm 1)$ depending on $w$ values (see below or table \ref{tab2}). Considering that $\arctanh\beta$ and $\arctanh\theta$ respectively tend to $\beta$ and $\mp 1/2\log\epsilon$ near the throat, we derive that there $\beta'\rightarrow 1+1/w$, $\mu\rightarrow -1/w$ and $P\rightarrow -1$ if we also assume that $\beta\log \epsilon$ is vanishing. This assumption is numerically checked by wormhole solutions presented below. It follows, as is well known, that the weak energy condition cannot be violated in $(\beta,\theta)=(0,\pm 1)$ when $-1<w<0$ since then $\mu>0$ and $w$ is not smaller than $-1$ (or from a geometrical viewpoint, the flaring-out condition $\beta'(r_0)>0$ is not respected). There is thus no wormhole for this range of values of $w$.
\begin{figure}[t]
\centering
\includegraphics[width=6cm]{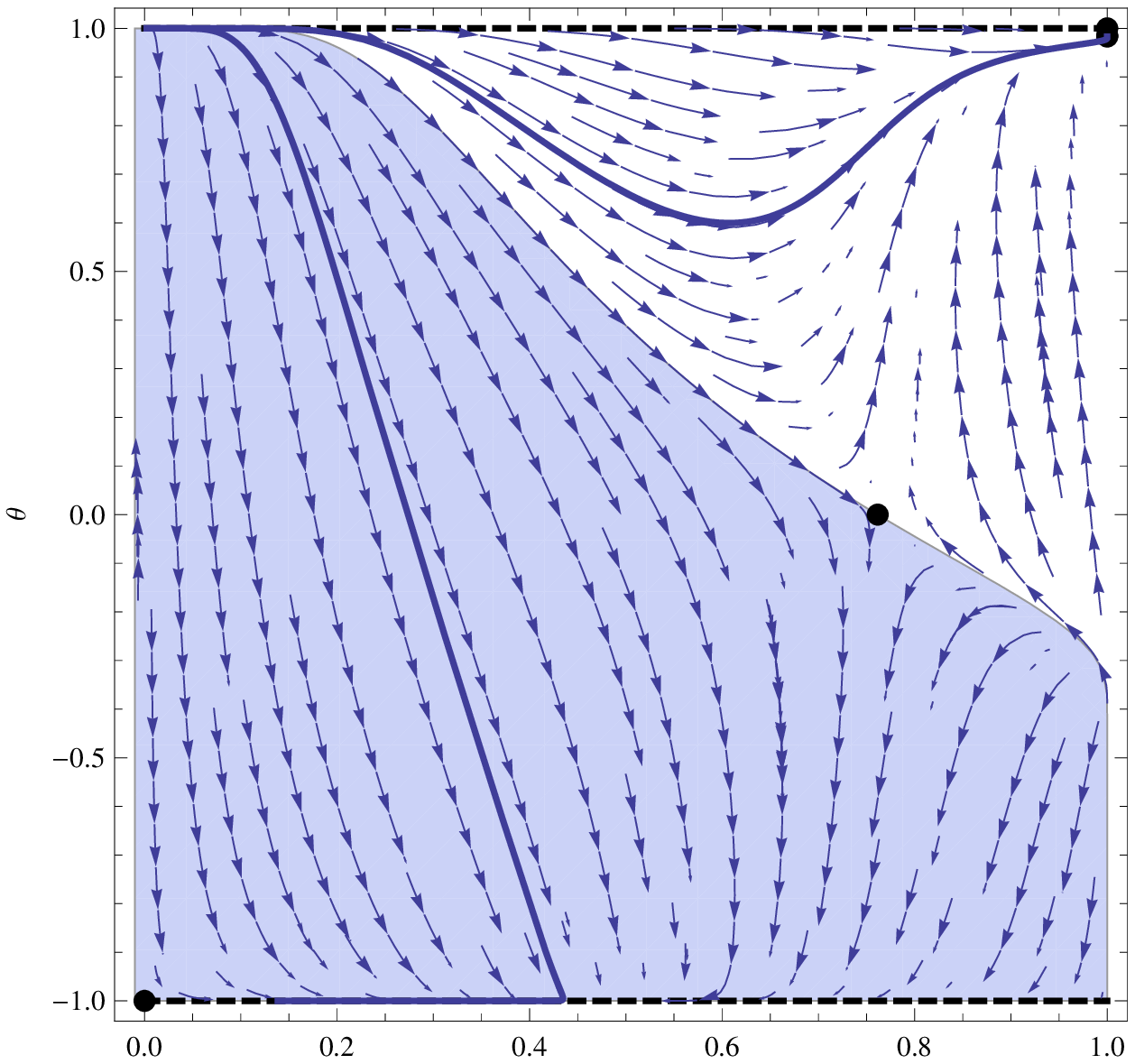}
\includegraphics[width=6cm]{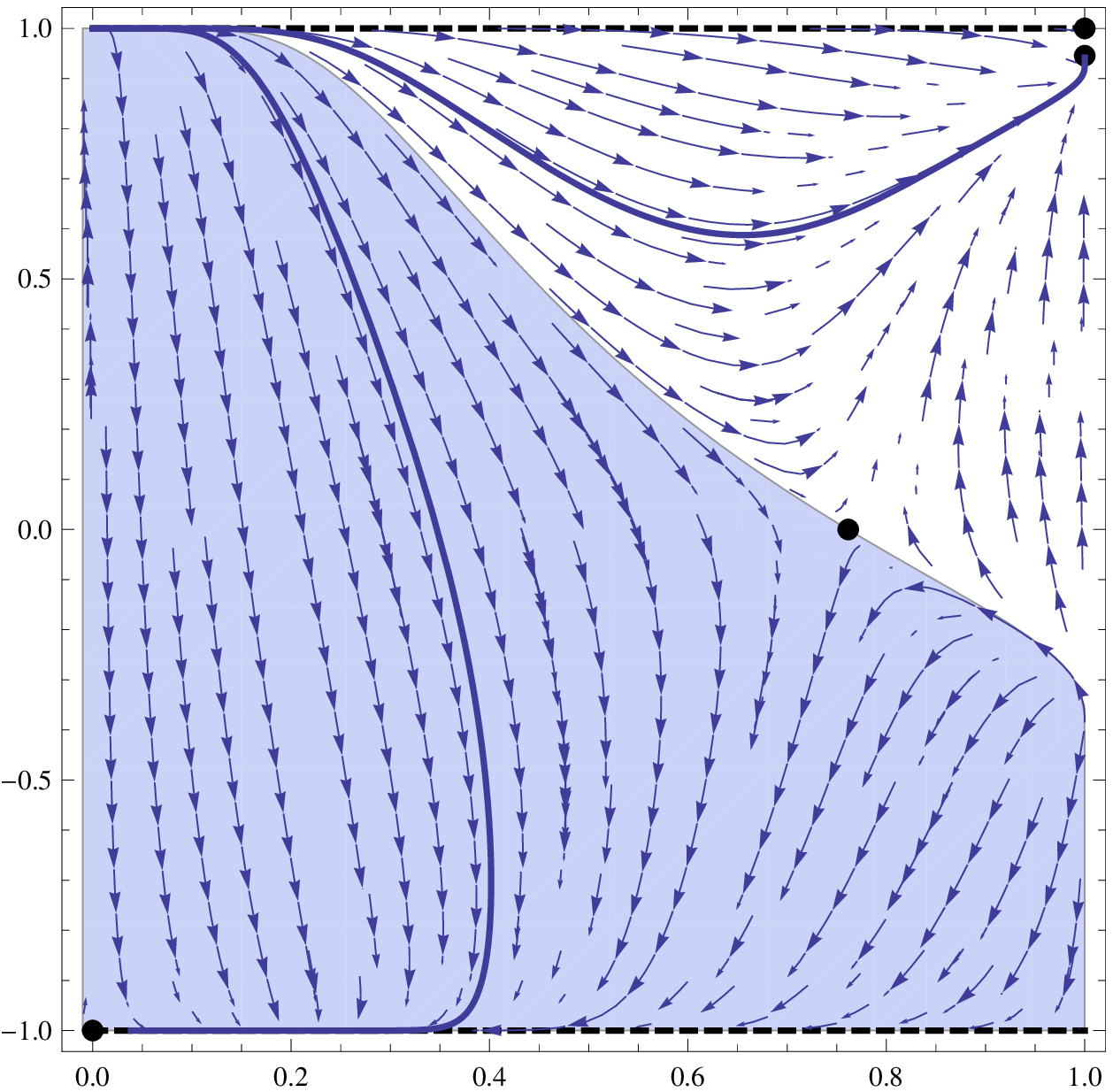}
\includegraphics[width=6cm]{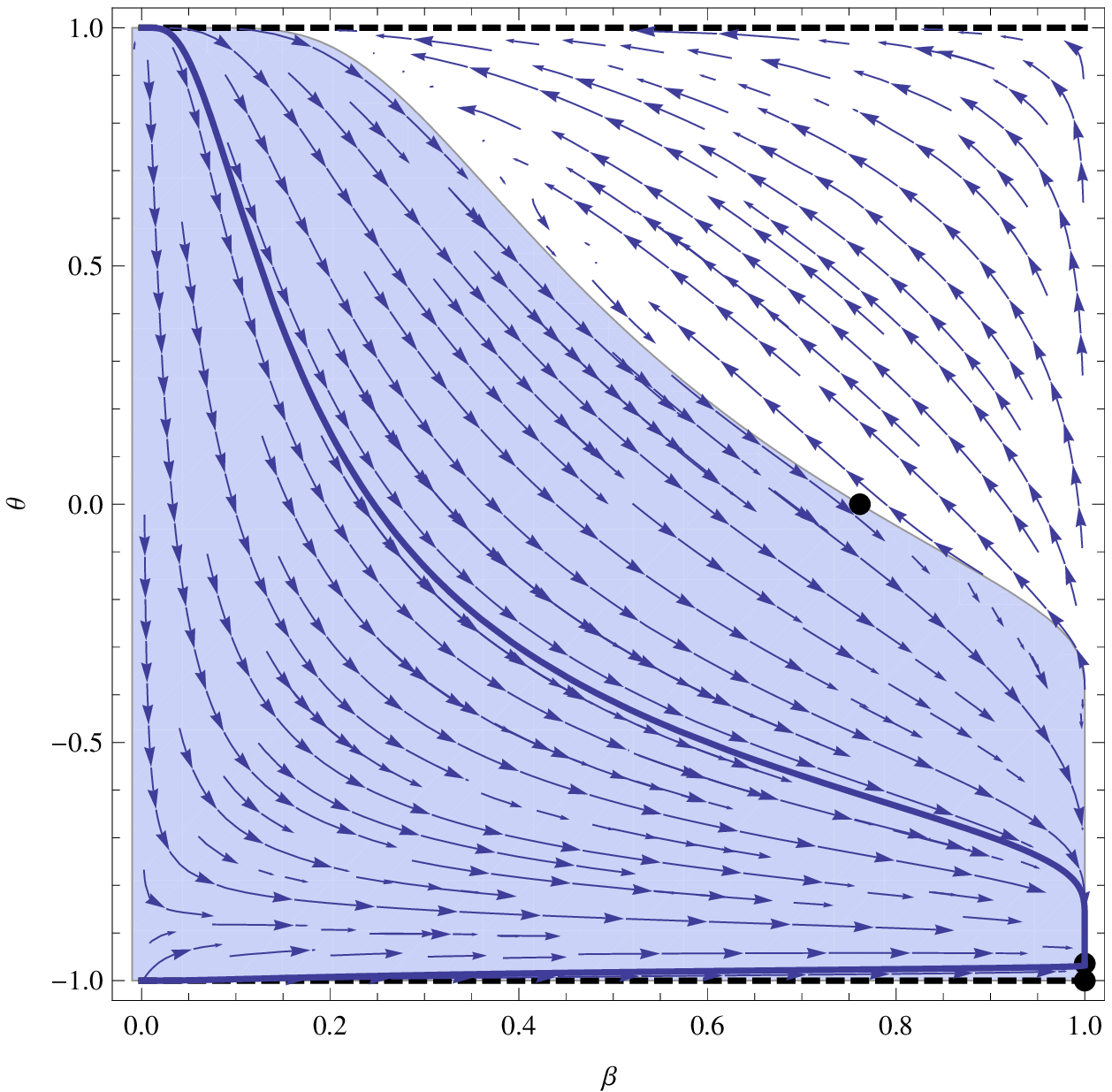}
\includegraphics[width=6cm]{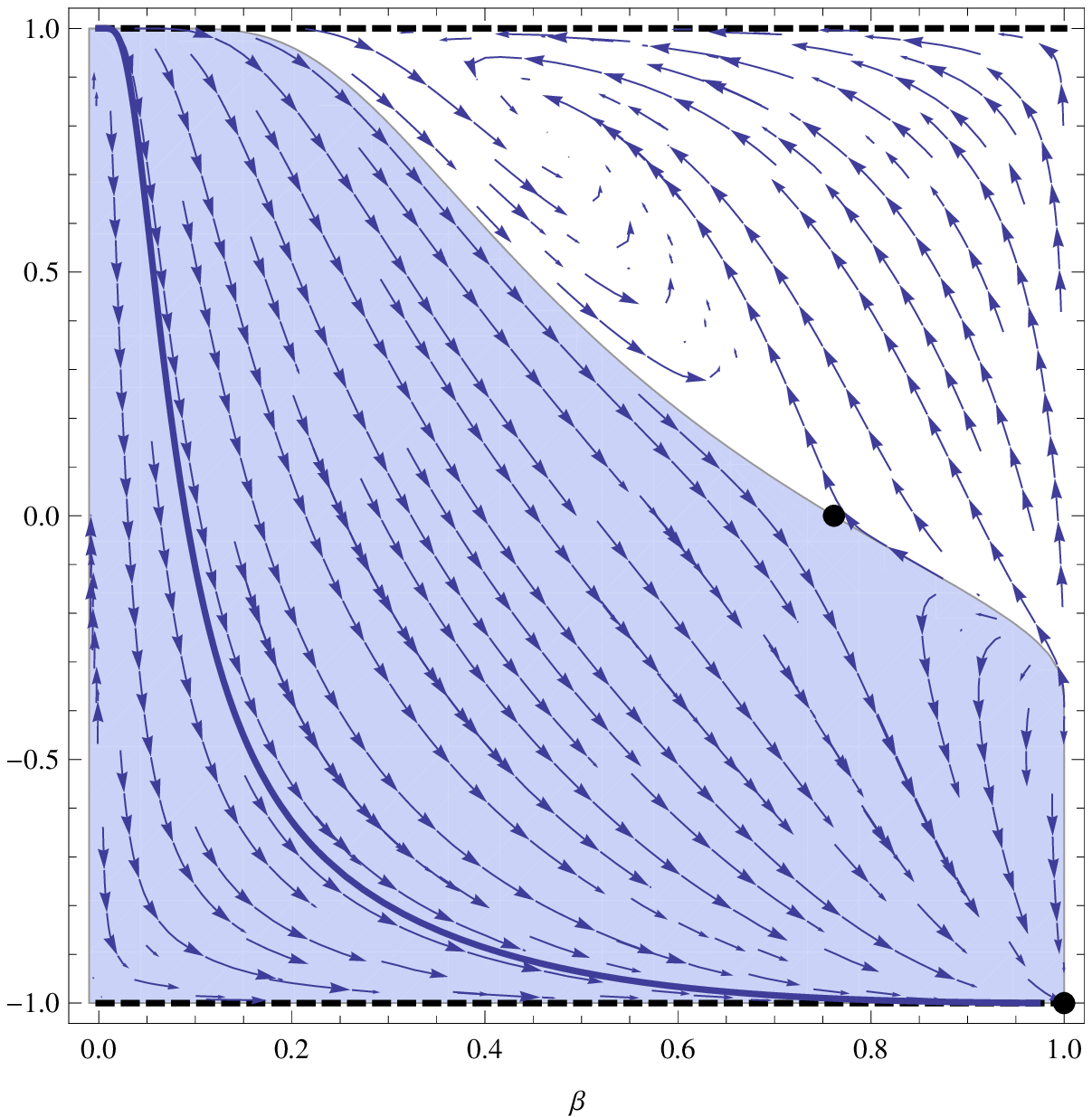}
\caption{\scriptsize{\label{fig3}Phase space of wormholes defined by a constant equation of state $w=-6.2$, $w=-2.3$, $w=0.2$ and $w=1.2$ (respectively the first, second, third and fourth graph). Thick trajectories and dots represent respectively some wormhole trajectories and their equilibrium points.}}
\end{figure}
\begin{table}
\begin{center}
\begin{tabular}{|l|c|c|c|c|c|}
\hline
{\scriptsize{$(\beta,\theta)$}} & \multicolumn{2}{c|}{\scriptsize{$-3-2\sqrt{2}$}} &\scriptsize{$-1...0$}& \multicolumn{2}{c|}{\scriptsize{$1$}} \\
\hline
$(\tanh 1,0)$ & \textbf{sa} & \textbf{sa} && \textbf{sa} & \textbf{sa} \\
$(\tanh\frac{(1+w)^2}{w^2+6w+1},\tanh\frac{2w}{1+w})$ & \textbf{si} & NE && si & si \\
$(1,-1)$ & sa & sa && \textbf{sa} & \textbf{si} \\
$(1,1)$ & \textbf{sa} & \textbf{sa} && sa & so \\
$(0,-1)$ & \textbf{si} & \textbf{si} && NE & NE \\
$(0,1)$ & NE & NE && sa & sa \\
$(1,\tanh(-1/2))$ & so & so && so & so \\
$(1,\tanh\frac{1+3w}{w-1})$ & \textbf{sa} & \textbf{si} && \textbf{si} & sa \\
\hline
\end{tabular}
\caption{\scriptsize{\label{tab1}This table shows the $8$ equilibrium points of the phase space for a constant equation of state such as $w\not\in\left[-1,0^-\right]$. "sa", "si" and "so" stand for "saddle", "sink" and "source". "NE" stands for "No Equilibrium". Those in bold belong to wormhole trajectories.}}
\label{label}
\end{center}
\end{table}
Consequently, we look for the phase space equilibrium points and their stabilities when $w\not\in\left[-1,0\right]$. They are summarised in table \ref{tab1}. All these points do not necessarily belong to trajectories describing wormholes. However, using numerical calculations, they allow to find the interesting intervals of $w$ that define the various families of wormhole trajectories. We then get
\begin{itemize}
\item When $w<-3-2\sqrt{2}$, a wormhole throat is in $(\beta,\theta)=(0,1)$. The trajectories end in $u\rightarrow \infty$, either in $(\beta,\theta)=(0,-1)$ and such as $P<0$ or in $(\beta,\theta)=(\tanh\frac{(1+w)^2}{w^2+6w+1},\tanh\frac{2w}{1+w})$ with $P>0$.
\item When $-3-2\sqrt{2}<w<-1$, a wormhole throat is in $(\beta,\theta)=(0,1)$. We recover the previous trajectories that end asymptotically in $(\beta,\theta)=(0,-1)$ such as $P<0$. A new set of trajectories ends asymptotically in $(\beta,\theta)=(1,\tanh\frac{1+3w}{w-1})$ with $P>0$.
\item When $0<w<1$, some wormholes trajectories start with a throat in $(\beta,\theta)=(0,1)$ and end in $(\beta,\theta)=(1,\tanh\frac{1+3w}{w-1})$, like above but with $P<0$ whatever $u$. A second set of wormhole trajectories starts in $(\beta,\theta)=(0,-1)$ and also ends asymptotically in $(\beta,\theta)=(1,\tanh\frac{1+3w}{w-1})$.
\item When $1<w$, wormholes trajectories have their throat in $(\beta,\theta)=(0,1)$ and end asymptotically in $(\beta,\theta)=(1,-1)$.
\end{itemize}
All these wormhole trajectories are plotted on figure \ref{fig3}. We checked numerically that at a throat, $P\rightarrow -1$ and $\beta'\rightarrow 1+1/w$. None of them tends to the Minkowski spacetime\cite{Lob05} since it is a saddle point whatever $w$, as indicated in table \ref{label}. Concerning the presence of singularities, since $P=-1$ at the throat, $\arctanh{\beta}\arctanh{\theta}$ is vanishing and there is none there. All the trajectories end in $u\rightarrow \infty$, without reaching the lines $\theta=\pm 1$ or $\beta=\pm1$ at finite $u$. The density and the scalars are thus finite everywhere at finite $u$ and these wormhole solutions are free from singularity.
%------------------------------------------------------------------------------%
\subsubsection{Equation of state as a redshift function}\label{s45}
\begin{figure}[t]
\centering
\includegraphics[width=7cm]{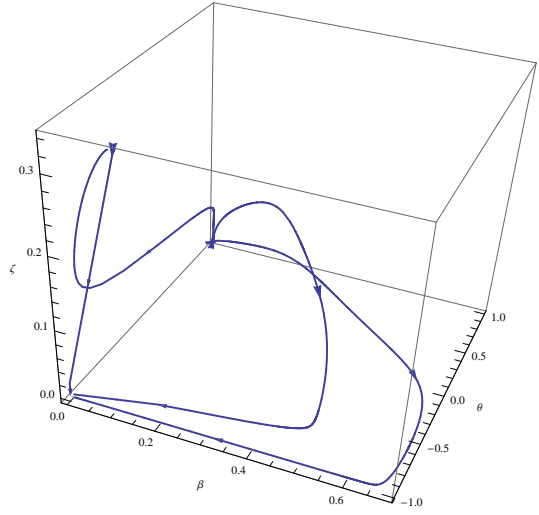}
\caption{\scriptsize{\label{fig14}Wormhole phase space for the cosmological equation of state $w=w_0+w_1 z$ with $w_0=-1.1$ and $w_1=-0.2$.}}
\end{figure}

Two equations of state are widely used in cosmology
\begin{itemize}
\item a linear equation of state $w=w_0+w_1 z$
\item the Chevallier-Polarski-Linder parameterisation $w=w_0+\frac{w_1 z}{1+z}$
\end{itemize}
$z$ being the redshift. To study the wormholes solutions of these dark energy, we look for a relation of the form $\mu=f(\mu,\beta,\theta,\zeta)$. For the linear equation of state, the conservation of the energy density for a FLRW cosmology gives
$$
\rho=3 e^{-3 w_1} H_0^2 \Omega_0 e^{3w_1(1+z)} (1+z)^{3+3 w_0-3 w_1}
$$
with $H_0$ and $\Omega_0$ the Hubble constant and the dark matter energy parameter. The equation of state being invertible, we get $z(w)=z(P/\mu)$ and
$$
\rho=3 e^{-3 w_1} H_0^2 \Omega_0 e^{3 (-w_0+w_1+\frac{P}{\mu })} (1+\frac{P-w_0\mu }{w_1\mu })^{3 (1+w_0-w_1)}
$$
Multiplying this last expression by $r^2$, we find
$$
\mu=3 e^{-3 w_1} H_0^2 \Omega_0 \frac{e^{3 (-w_0+w_1+\frac{-1+\arctanh{\beta}(1+2 \arctanh{\theta})}{\mu })} \arctanh{\beta} (1+\frac{-1-w_0 \mu +\arctanh{\beta}(1+2 \arctanh{\theta})}{w_1 \mu })^{3 (1+w_0-w_1)}}{\arctanh{\zeta}^2}
$$
With this expression, it is now possible to look for wormholes in the phase space $(\beta,\theta,\zeta)$. Since $\mu(\zeta)=\mu(-\zeta)$, their trajectories are symmetric with respect to the plane $\zeta=0$. We choose to consider a dark energy with $-1.5<w_0<-1,-0.5<w_1<0$ and cosmological parameters $H_0=70$, $\Omega_{m0}=0.27$ in agreement with the observations\cite{Rud15,Jas10,Tsu09}. We get the figure \ref{fig14}. All the trajectories have their throat in $(\beta,\theta,\zeta)=(0,1,0)$ and end in $(\beta,\theta,\zeta)=(0,-1,0)$ with diverging $l$ and density. Hence, they do not have any singularity and are not asymptotically Minkowski flat.\\
The same calculations can be made with the Chevallier-Polarski-Linder parameterisation with then 
$$
\mu=3 e^{-3 w_1} H_0^2 \Omega_0 \frac{e^{3 \left(w_0+w_1-\frac{-1+\arctanh{\beta}(1+2 \arctanh{\theta})}{\mu }\right)} w_1^3 \mu ^3 \arctanh{\beta} \left(\frac{w_1 \mu }{1+\mu(w_0+w_1)-\arctanh{\beta} (1+2 \arctanh{\theta})}\right)^{3 (w_0+w_1)}
}{\arctanh{\zeta}^2 (1+\mu(w_0+w_1)-\arctanh{\beta} (1+2 \arctanh{\theta}))^3}
$$
and the same ranges for the parameters $w_0$ and $w_1$. Once again, $\mu(\zeta)=\mu(-\zeta)$. We get a figure similar to figure \ref{fig14}. The linear and Chevallier-Polarski-Linder dark energy thus lead to the same type of wormholes solutions.
%------------------------------------------------------------------------------%
\subsubsection{Navarro-Frenk-White}\label{s44}
\begin{figure}[h]
\centering
\includegraphics[width=8cm]{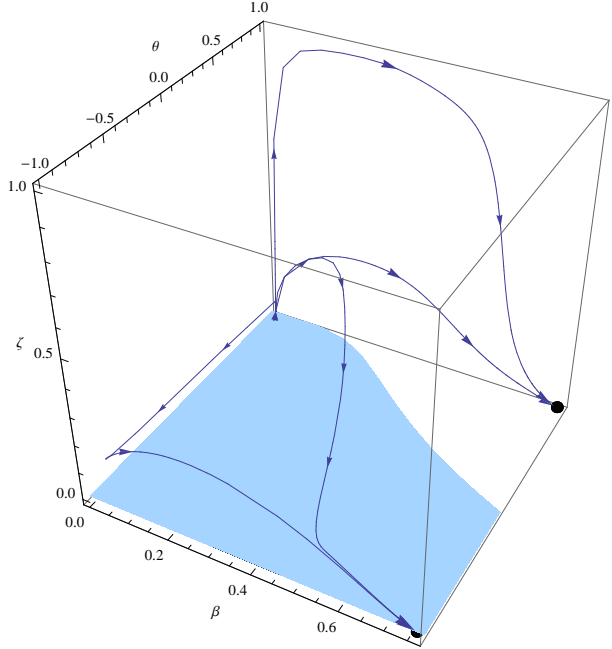}
\caption{\scriptsize{\label{fig9}The two families of wormhole trajectories for a Navarro-Frenk-White distribution of matter when $\rho_s=1$ and $r_s=1$. Each curve is symmetric with respect to the $\zeta=0$ plane. The two wormhole families are separated by the sign of $P$ which is projected on the plane $\zeta=0$ but near the throat where they both are such as $P=-1$.}}
\end{figure}

A Navarro-Frenk-White profile\cite{Nav96} is generally used to fit dark matter halos. It has also been shown \cite{Rah14} that a galactic halo with such a profile and observed rotation curves has the properties for supporting traversable wormhole sustained by this fluid in the anisotropic case $p\not = \tau$. Also the NFW profile is now considered in the wormhole literature\cite{Isl18, Rah16}. In this subsection, we determine the different families of wormhole solutions for this dark matter in the isotropic case, showing that none of them is asymptotically Minkowski flat. The NFW density writes as
$$
\rho=\frac{\rho_s}{\frac{r}{r_s}(1+\frac{r}{r_s})^2}
$$
with $\rho_s$ and $r_s$ some positive constants. Using the definition (\ref{var6}), we get
$$
\mu= \frac{\pm r_s^3 \rho_s \sqrt{\arctanh\beta} \arctanh\zeta}{(\pm \sqrt{\arctanh\beta}+r_s \arctanh\zeta)^2}
$$
with $\pm$ corresponding to the upper (lower) part of the wormhole. Once again $\mu(\zeta) =\mu(-\zeta)$ and we limit the phase space study to $0<\zeta<1$, the phase space trajectories being symmetric with respect to the plane $\zeta=0$. At the throat, it is easy to show that $\beta'>0$ if $1-\frac{\pm r_0r_s^3\rho_s}{\pm r_0+r_s}^2>0$. Moreover, when $\beta=\tanh 1$, $\beta'<0$ indicating that the wormhole trajectories coming from the throat are such as $\beta<\tanh 1$. The equilibrium points respecting these limits for $\beta$ and $\zeta$ and intervening in the wormholes trajectories are:
\begin{itemize}
\item $(\beta,\theta,\zeta)=(\tanh 1,\tanh 2,0)$ which is a sink when $\zeta\rightarrow 0^+$
\item $(\beta,\theta,\zeta)=(\tanh 1,-1,0)$ which is a sink when $\zeta\rightarrow 0^+$ 
\end{itemize}
We have then two families of wormhole trajectories, symmetric with respect to the $\zeta=0$ plane. They are separated by the surface $P=0$ for $0<\zeta<1$ (that is such as $\arctanh\beta=1-\mu$ and thus $\beta'=0$) but in the neighbourhood of the throat in $(\beta,\theta,\zeta)=(0,1,0)$ where $P=-1$. The first family ends in $(\beta,\theta,\zeta)=(\tanh 1,\tanh 2,0^+)$ with $P>0$ and the second one in $(\beta,\theta,\zeta)=(\tanh 1,-1,0^+)$ with $P<0$, both when $r\rightarrow +\infty$. These non singular trajectories thus tend to a spatially flat metric but not to a flat Minkowski spacetime. Some examples of these curves are plotted on figure \ref{fig9}.
%------------------------------------------------------------------------------%
\subsection{A toy model of asymptotically flat wormhole on both side of its throat}\label{s42}
In this last subsection, we consider a toy model of static wormhole that is asymptotically Minkowski flat\footnote{We do not consider the Bronnikov-Ellis model\cite{Bro73,Ell73} which is not based on an isotropic perfect fluid.}. As a physical result, we show that such a wormhole can be asymptotically flat on both sides of its throat. This is thus a counterexample to a recent theorem\cite{Bro17} that excludes such a possibility for a static wormhole sustained by an isotropic fluid.\\\\
We consider the toy model defining by the following pressure
$$
p=p_0r^{-n}
$$
with $n>2$ since then, when $r$ is diverging, the pressure vanishes faster than $r^{-2}$ as required to get a Minkowski equilibrium point as a sink (see subsection \ref{s21}). Such a model has already been considered in \cite{Lob13,Rah07}. From equation (\ref{eq3A}) and the fact that $d\Phi/dr=e^{-u}\arctanh\theta $, we get
$$
w=p/\rho=P/\mu=\frac{\arctanh\theta}{n-\arctanh\theta}
$$
This equation of state is independent from $p_0$ and tends to $-1^-$ in $\theta\rightarrow +1$. The weak energy condition is thus violated that explains the presence of a throat in $(\beta,\theta)=(0,1)$. Then using (\ref{eq3}), it comes that
$$
\mu= \frac{(n-\arctanh\theta)(-1+\arctanh\beta+2\arctanh\beta\arctanh\theta)}{\arctanh\theta}
$$
$\mu$ is independent on $\zeta$ and the dynamical system thus reduces to two differential equations for $\beta'$ and $\theta'$. The flaring-out condition $\beta'(r_0)>0$ implies $p_0r_0^{2-n}>-1$.
\begin{figure}[!htb]
\centering
\includegraphics[width=6cm]{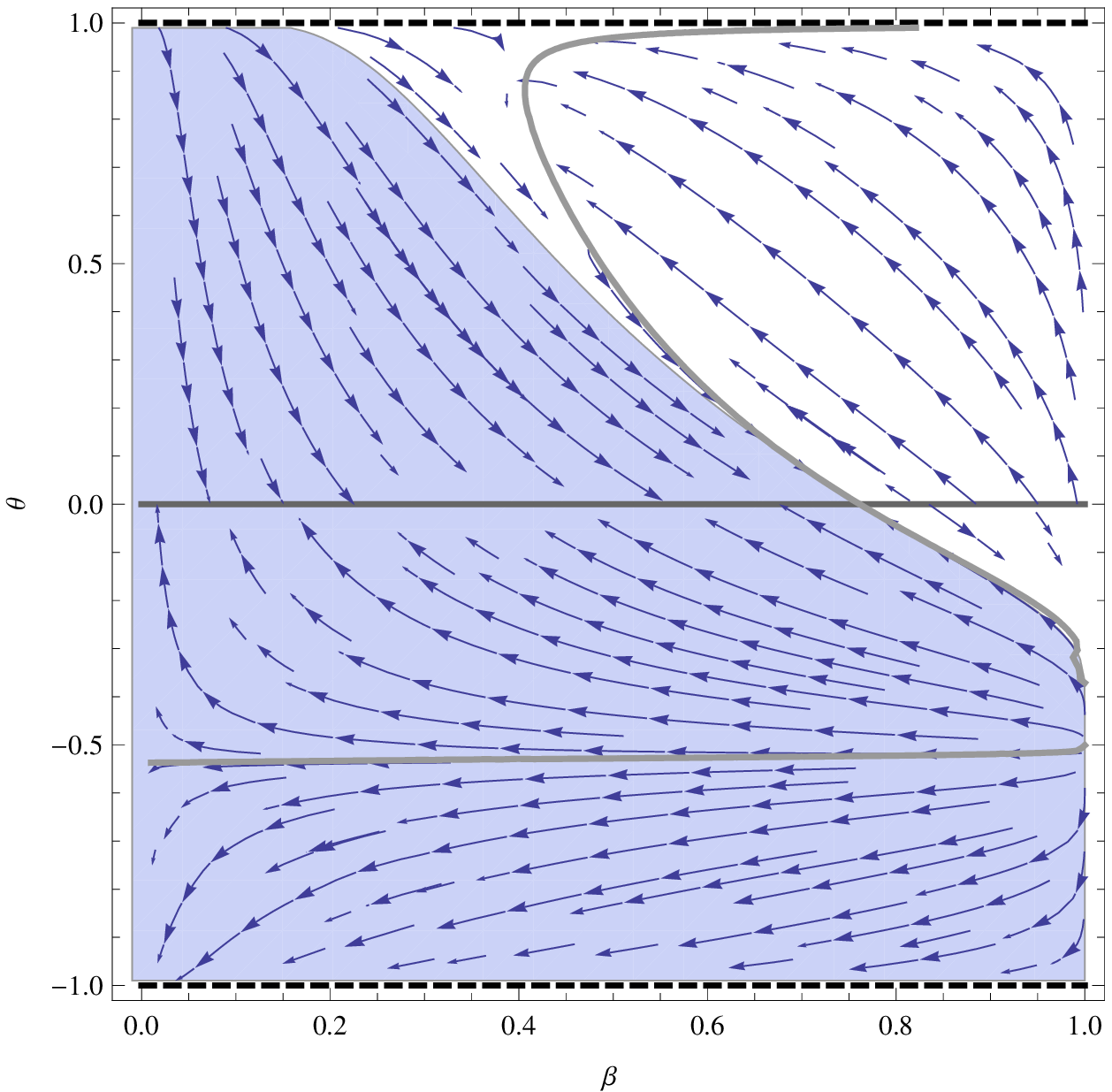}
\includegraphics[width=6cm]{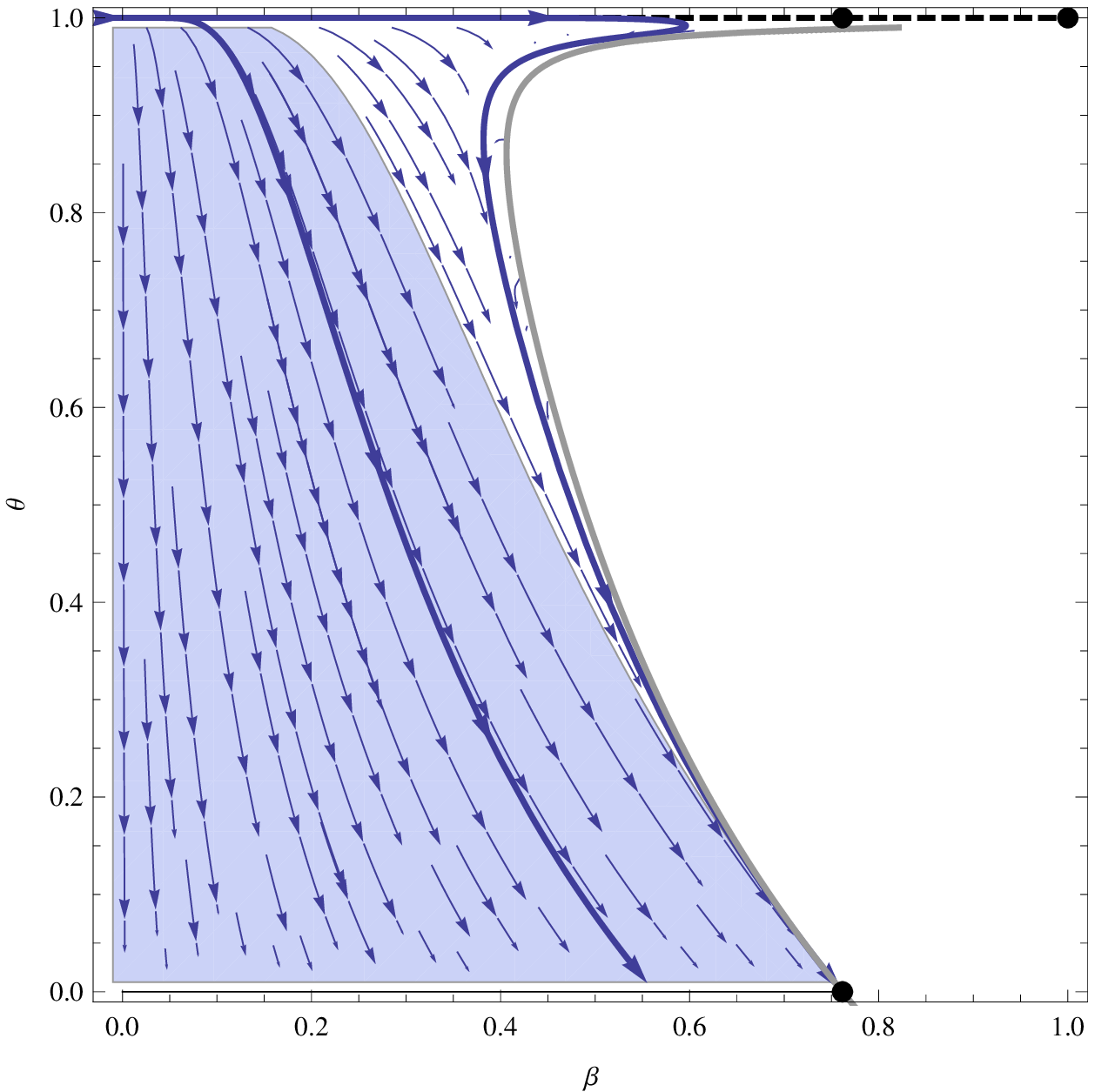}
\includegraphics[width=6cm]{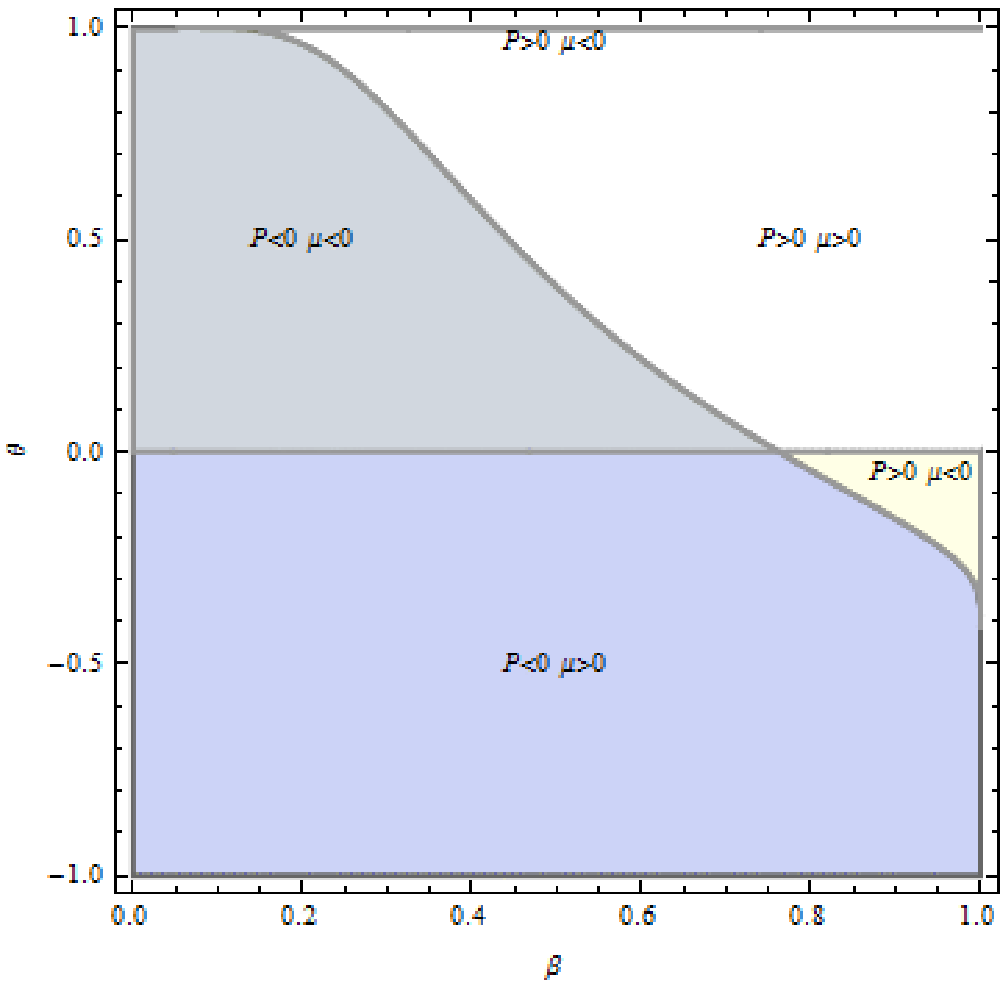}
\includegraphics[width=6cm]{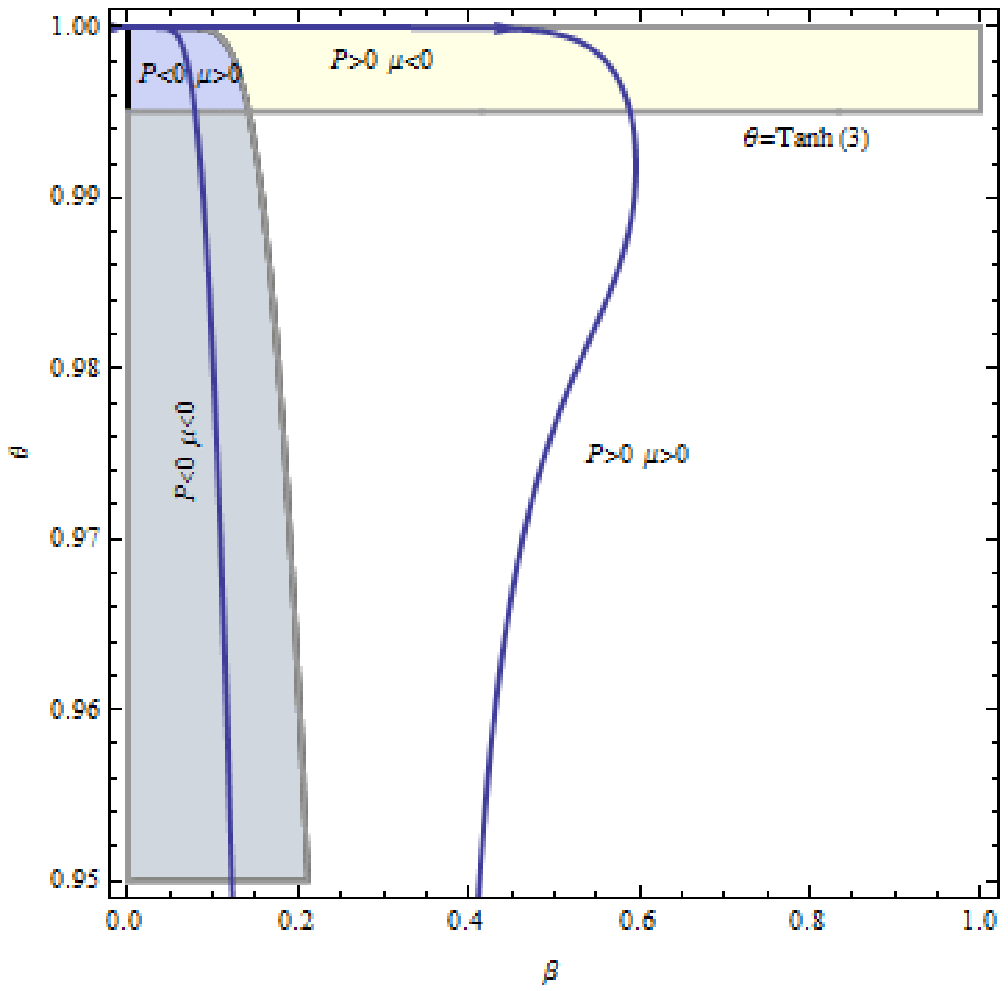}
\caption{\scriptsize{\label{fig4}Phase space representations for $n=3$. The first graph shows the different parts of the phase space, separated by some gray curves corresponding to $\theta=0$ and $\theta'=0$. Only the part on the upper left corner contains wormhole trajectories which are shown on the second graph with the three equilibrium points. Two numerical wormhole solutions are plotted for $(r_0,p_0)=(0.1,-0.5)$ and $(r_0,p_0)=(0.02,0.1)$ (black bold curves, the right one ending at the Minkowski equilibrium point). The third and fourth graphs show the signs of $P$ and $\mu$ in the phase space. The fourth graph is an enlargement of the third one for $\theta\simeq 1$ and shows parts of the numerical solutions of the second graph}}
\end{figure}
We then get the phase space trajectories plotted on the first graph of figure \ref{fig4} for $n=3$ (similar figures are obtained for other values of $n$). The line $\theta=0$ and the curve $\theta'=0$ (in gray on the mentioned first graph) split the trajectories in four sets. Only the one with $\theta>0$ and $\theta'<0$, shown on the second graph of figure \ref{fig4}, contains the wormhole trajectories. Some numerical solutions are plotted in bold (with enlargement for $\theta\simeq 1$ on the fourth graph). The signs of $\mu$ and $P$ are also shown on the third and fourth graphs on figure \ref{fig4}. Equilibrium points of wormholes trajectories are the following
\begin{itemize}
\item $(\beta,\theta)=(\tanh 1,0)$. This is the Minkowski equilibrium point whose stability is confirmed by numerical simulations in $\zeta=0^+$ for $P>0$.
\item $(\beta,\theta)=(\tanh 1,1)$ is a saddle.
\item $(\beta,\theta)=(1,1)$ is also a saddle.
\end{itemize}
We have then two sets of phase space trajectories representing wormhole solutions, that both have their throats in $(\beta,\theta)=(0,1)$
\begin{itemize}
\item For the first set, a trajectory starts with a negative pressure (i.e. $p_0<0$) and positive density and ends with a singularity on the line $\theta=0$ with $\beta<\tanh 1$. At the singularity, the density and the scalars diverge whereas the pressure is finite. We checked the presence of this singularity with the geodesic equation for $u(\tau)$ whose solution for a wormhole trajectory defined by $r_0=1$ and $p_0=-0.5$ is plotted on figure \ref{fig7}, showing that $u$ and $\tau$ end when $\theta$ reaches zero.
\item For the second set, a trajectory starts with a positive pressure (i.e. $p_0>0$) and negative density and it goes asymptotically to the Minkowski equilibrium point where the pressure, the density and $w$ tend to vanish. We thus recover an empty Universe far from the throat. Such a trajectories is plotted in the white area of the second graph on figure \ref{fig4}. Numerical simulations show that the larger $n$, the smaller the area where stand these trajectories.
\end{itemize}
Since $\mu$ does not depend on $\zeta$, we should thus have a flat spacetime on both sides of the wormhole. Let us show that this is a counterexample to the no-go theorem of \cite{Bro17} that excludes such a behaviour of the metric in presence of an isotropic perfect fluid. To show this no-go theorem, the authors used the form of the metric $ds^2=A(x)dt^2-dx^2/A(x)-r^2(x)d\Omega^2$. The no-go theorem then rests upon the fact that the quantity $D=A r^2>0$ should go to $+\infty$ on both sides of the throat when spacetime becomes flat with $A\rightarrow 1$ and $r\rightarrow +\infty$. $D$ should thus possess a minimum at the throat such as $dD/dx=0$. But the behaviour of $D$ is given by the equation (14) in \cite{Bro17} that writes
$$
\frac{d^2D}{dx^2}-4\frac{1}{r}\frac{dD}{dx}\frac{dr}{dx}+4\frac{D}{r^2}(\frac{dr}{dx})^2+2=0
$$
Hence, when $dD/dx=0$, $d^2D/dx^2<-2$, that implies a maximum for $D$ and not a minimum, thus leading to the no-go theorem. Let us show that the model of the present subsection is a counterexample to this theorem. Adopting the same metric signature, we have $D=e^{2\Phi}r^2$ and $dr/dx=\mp\sqrt{\arctanh\beta}e^{-\Phi}$. $\mp$ stands for above or below the throat where $dr/dx$ vanishes, $r$ reaching the minimum value $r_0$. We also have that $P\rightarrow p_0r_0^{2-n}$ and thus $\arctanh\beta\arctanh\theta\rightarrow 1/2(p_0r_0^{2-n}+1)$. Consequently, $\frac{dD}{dx}\frac{dr}{dx}=2r(\arctanh\theta+1)\arctanh\beta$ tends to the non vanishing constant $r_0(p_0r_0^{2-n}+1)$ at the throat and $\frac{dD}{dx}$ tends to $\mp\infty$ on both sides of it, $D$ being a finite quantity (since in $r\rightarrow r_0^+$, $\theta\rightarrow 1$, $\Phi'\rightarrow +\infty$ and thus $\Phi$ cannot diverge positively). An extremum value of $D(x)$ is thus reached at the throat but with $dD/dx\not =0$, a case not taken into account in \cite{Bro17}. This extremum is also a minimum since $4\frac{D}{r^2}(\frac{dr}{dx})^2=0$, $p_0$ is positive for the trajectories ending at the Minkowski sink point and thus $\frac{d^2D}{dx^2}=2+4p_0r_0^{2-n}>0$. Hence, the no-go theorem of \cite{Bro17} does not apply to the model of this subsection and we have an asymptotic Minkowski flat spacetime on both sides of the throat.
\begin{figure}[h]
\centering
\includegraphics[width=8cm]{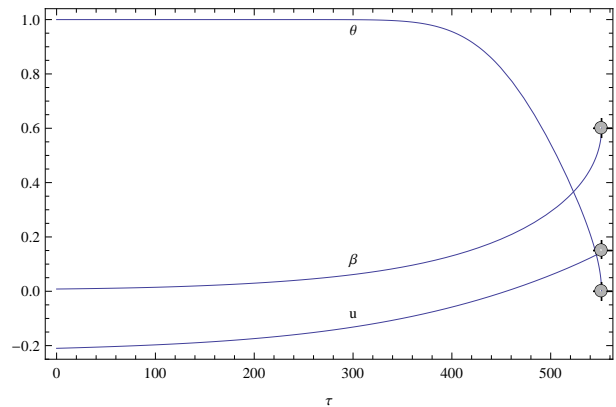}
\caption{\scriptsize{\label{fig7}$u(\tau)$, $\beta(\tau)$ and $\theta(\tau)$ for a wormhole trajectory defined by $n=3$, $r_0=1$ and $p_0=-0.5$. This trajectory is plotted in bold in the phase space on figure \ref{fig4}. The gray points indicate the ends of the curves $(u,\beta,\theta)$ in $\tau=551$ and $u=0.14$.}}
\end{figure}
%------------------------------------------------------------------------------%
\section{Conclusion}\label{s5}
In this paper we rewrote the field equations of static spherically symmetric wormholes sustained by an isotropic perfect fluid as a dynamical system with normalised variables. It allows displaying at a glance all the solutions of the field equations in a finite phase space without having to solve them. It offers a unified framework to compare different wormhole models and to study their global properties. We determined the possible locations of throats, equilibrium points as well as curvature singularities in the phase space. We also associated a geodesic equation to spacetime that describes a phase space trajectory. The wormhole throats stand in $(\beta,\theta,\zeta)=(0,\pm 1,0)$. They are not equilibrium points despite all the wormhole trajectories pass through them. Their location allow to prove a first general physical result, i.e. that the variation of the gravitational redshift at the throat of a static and isotropic wormhole is always diverging (but $\Phi$ can be finite), generalising the recent result that there is no zero-tidal force static spherically symmetric wormhole sustained by an isotropic perfect fluid. The equilibrium point corresponding to the Minkowski solution is in $(\beta,\theta,\zeta)=(\tanh1,0,0)$ and can only be reached by solutions whose pressure and density of the fluid decrease faster than $r^{-2}$. We then used this formalism to study the dynamical systems of several static spherically symmetric wormholes models, getting other physical results.\\\\
The first model we studied is defined by a vanishing density $\rho=0$. From a mathematical viewpoint, its exact solution allows checking the agreement of our dynamical system of equations with the usual wormhole field equations (\ref{eq1A}-\ref{eq3A}). This model has only one family of wormhole solutions in the two dimensional phase space $(\beta,\theta)$. Its trajectories reach a singularity at finite distance and time from the throat. From a physical viewpoint, this model contains as a special case ($P=0$) the Schwarzschild black hole. We have plotted it in the phase space to show what looks like its trajectory. It is discontinued at the horizon in $(\beta,\theta,\zeta)=(0,\pm 1,0)$, reflecting the discontinuity of the Schwarzschild metric in $r=2M$ with $(r,t)$ coordinates that are not appropriate to study horizons. This is the only type of black hole such a fluid admits. Since this paper is dedicated to static spherically symmetric wormholes, we did not pursue further in the study of static spherically symmetric black holes, nor horizons that would need a different set of coordinates to be clearly identified in the phase space $(\beta,\theta,\zeta)$. However, this shows that the dynamical system analysis of this paper could also be used to consider these black holes and their horizons.\\ 
We then look for some wormholes that could be generated by some usual dark energy and dark matter models. Hence, the generalised Chaplygin gas has wormhole solutions when $A<r_0^{-2(1+\alpha)}$. There is only one family of trajectories describing identical wormholes connected the ones with the others and thus repeating cyclically.\\
A constant equation of state $p/\rho=w$ has a richer dynamics. It has five families of wormhole solutions. As is well known, we have no wormhole for $-1<w<0$ since then the weak energy condition cannot be violated. Wormholes exist for a ghost dark energy with $w<-1$ and a positive density or for $w>0$ with a negative density at the throat. All the trajectories end asymptotically with a diverging $r$ at some equilibrium points. None of them possesses a singularity (at finite $r$). Some of them are asymptotically spatially flat but do not tend to a Minkowski spacetime.\\
We also look for the wormhole solutions that can be generated by the linear and CPL equations of state (depending on the redshift $z$) widely used in the literature to study dark energy. Taking the range of their parameters in agreement with observations and a ghost dark energy ($w<-1$), we found that they possess only one family of solutions, describing some wormholes without singularity at finite $l$ but that are not asymptotically flat.\\
Last, the static wormhole model defined by a Navarro-Frenck-White density profile has wormhole solutions when $1-\frac{\pm r_0r_s^3\rho_s}{\pm r_0+r_s}^2>0$. Two wormhole families are found with trajectories that both end in $r\rightarrow +\infty$ with a spatially flat spacetime different from the Minkowski one.\\
The main physical result about these five dark energy and dark matter models that are often used to describe our Universe is that none of them can sustain a static wormhole that could be asymptotically Minkowski flat. This tends to show that they should not form naturally some static wormholes.\\\\
Finally, we consider a toy model such that the pressure vanishes as $r^{-n}$ with $n>2$. There are then two families of trajectories describing wormholes. For one of them, the trajectories end at a singularity at finite $r$ and time. For the other one, the trajectories asymptotically tend to the Minkowski equilibrium point far from the throat. The physical result here, is that this flatness occur for both sides of the wormhole. This is thus a counterexample to a theorem claiming that a static wormhole sustained by a perfect fluid could not be flat on both sides of its throat. This is interesting since it has not to be glued to a vacuum exterior spacetime to be physically acceptable\cite{Lem03,Hos18}.\\\\
Let us conclude with some possible extension of this work. Dynamical system equations (\ref{eq1}-\ref{eq4}) with the constraint on $T$ allow generalising the framework we develop to non-isotropic forms of perfect fluid. Hence, it could be interesting to study the cases when $p/\rho$ and $\tau/\rho$ are some constants for instance. The dynamical system could also be used to study relativistic star models defined by their barotropic equation of state as in \cite{Nil01,Nil01A,Hei03} or static black holes as shown with the Schwarzschild one.
%------------------------------------------------------------------------------%
\bibliographystyle{unsrt}

\end{document}